\documentclass[preprint]{aastex631}
\usepackage{amsmath}

\shorttitle{Interior temperature of Uranus and Neptune}
\shortauthors{B. Militzer}

\graphicspath{{./}{figures/}}

\newcommand{\rr}{\ensuremath{\vec{r}}}

\newcommand{\gcc}{g$\,$cm$^{-3}$}

\begin{document}
%\nolinenumbers

\title{Ab Initio Entropy Calculations of Water Predict the Interiors of Uranus and Neptune to be 15--30\% Colder than Previous Models}

\author{Burkhard Militzer}
\affil{Department of Earth and Planetary Science, Department of Astronomy,\\ University of
California, Berkeley, CA, 94720, USA}

\begin{abstract}\nolinenumbers
  {\it Ab initio} free energy calculations are employed to derive the
  entropy of liquid and superionic water over a wide range of
  conditions in the interiors of Uranus and Neptune. The resulting
  adiabats are much shallower in pressure-temperature space than those adopted for
  earlier models of Uranus and Neptune. Our models for their interiors
  are thus much colder, increasing the likelihood that diamond
  rain or the recently predicted phase separation of planetary
  ices has occurred in the mantles of ice giant planets. Based on our
  {\it ab initio} data, we construct interior models for Uranus and Neptune
  with the Concentric MacLaurin Spheroid method that match the
  existing gravity measurements. We compare fully convective models
  with models that include a convective boundary between liquid and
  superionic water. We also share a code to characterize giant planet
  atmospheres where para and ortho hydrogen as well as helium are
  present.
\end{abstract}

\section{Introduction} \label{sec:intro}

The primary mechanism for secular cooling within planets is convection
because heat transport via diffusion and radiation is comparatively
inefficient \citep{hubbard_planets}. The magnetic fields of Jupiter,
Saturn, Uranus and Neptune but also that of Earth and Mercury provide
direct evidence that convection occurs in their interiors. Ample
evidence of plate tectonics in our plant demonstrates that convection
also occurs in solid, highly viscous layers. For Uranus and Neptune this
implies that if there existed a layer of superionic water in their
interior, it would mostly likely be convective
\citep{matusalem2022plastic}. For convection not to occur either
requires an endothermic first order phase transition
\citep{schubert1975role} or the presence of a gradient in composition
\citep{leconte2012new}. If a planet's interior has differentiated
into multiple convective layers the most plausible assumption would be
to represent each layer by an adiabat and to match pressure and
temperature at the boundaries \citep{hubbard_planets}. The entropy of
the outermost layer can typically be constrained by observations.

NASA has made a Uranus orbiter and probe mission a priority for this
decade \citep{decadal} generating new interest in understanding the
formation and interior structure of Uranus and Neptune. Comparatively
little is known about these two ice giant planets because so far they
have only been visited by only a single spacecraft, Voyager 2, which
during its fly-bys in 1986 and 1989 made the surprising discovery that
both planets have nondipolar magnetic
fields. \citet{ruzmaikin1991origin} and
\citet{Stanley2004,Stanley2006} showed that such fields emerge if they
are generated from convection restricted to a thin outer mantle
layer. Consistent with that interpretation, \citet{militzer2024phase}
proposed the outer layer is composed of water and hydrogen while there
is an inner layer composed of carbon, nitrogen and a depth dependent
concentration of hydrogen, which stabilizes this layer against
convection. This structure was based on the spontaneous phase
separation of planetary ices that was observed in {\it ab initio}
simulations of H$_2$O, CH$_4$, and NH$_3$ at high pressure and
temperature. The resulting models were consistent with the gravity
measurements and the proposed geometry for the magnetic dynamo layer.

As an alternative explanation for the nondipolar fields,
\citet{soderlund2013turbulent} performed magneto-hydrodynamic
simulations of Uranus's and Neptune's magnetic fields assuming thick
and thin shell dynamos. They suggested that the nondipolar magnetic
fields are the result of turbulent convection that is driven by
thermal buoyancy. Various dynamo types are discussed by
\citet{soderlund2020}.

The interpretation of the gravity measurements has not been
unique~\citep{movshovitz2022promise} but it is generally assumed that
a large part of their interior is composed of planetary ices, H$_2$O,
CH$_4$, and NH$_3$ \citep{Podolak1995}. \citet{Helled2020} reviewed
the different methods and assumptions that have been invoked to
characterize the structure and evolution of Uranus and Neptune.
\citet{Nettelmann2013} matched the available gravity
data with interior models consisting of three layers, each homogeneous
and convective. The outer two layers are composed of hydrogen, helium,
and water, though their compositions differ, while the innermost layer
is a rocky core.

\citet{Helled_2014} developed core-accretion models for Uranus and
Neptune, investigating the conditions that resulted in the observed
masses and solid-to-gas ratios.
\citet{Bailey_2021} suggested that the difference in heat flux between
Uranus and Neptune indicates varying levels of water-hydrogen mixing
in their outer envelopes.
\citet{movshovitz2022promise} constructed ensembles of interior models
for Uranus and Neptune with agnostic pressure-density relationships,
constrained their moments of inertia and discussed gravity
measurements of a future low-periapse orbiter.

\citet{neuenschwander2024uranus} investigated the
possible relationships of pressure, density, temperature and
composition by incorporating convective and nonconvective layers into
Uranus's interior and compared models that include a water-rich layer
with those that do not.
Most recently \citet{lin2025interior} constructed interior models for
Uranus under a variety of assumptions and concluded that its mantle 
cannot be composed solely of water. \citet{french2024uranus} provided
additional constraints for Uranus's gravity coefficients by analyzing
occultation data for its rings.
\citet{stixrude2021thermal} investigated the thermal evolution of
Uranus's interior and proposed that the lack of a strong heat flux
might be due to the formation of a growing core composed of superionic
water.

\begin{figure}[ht!]
\plotone{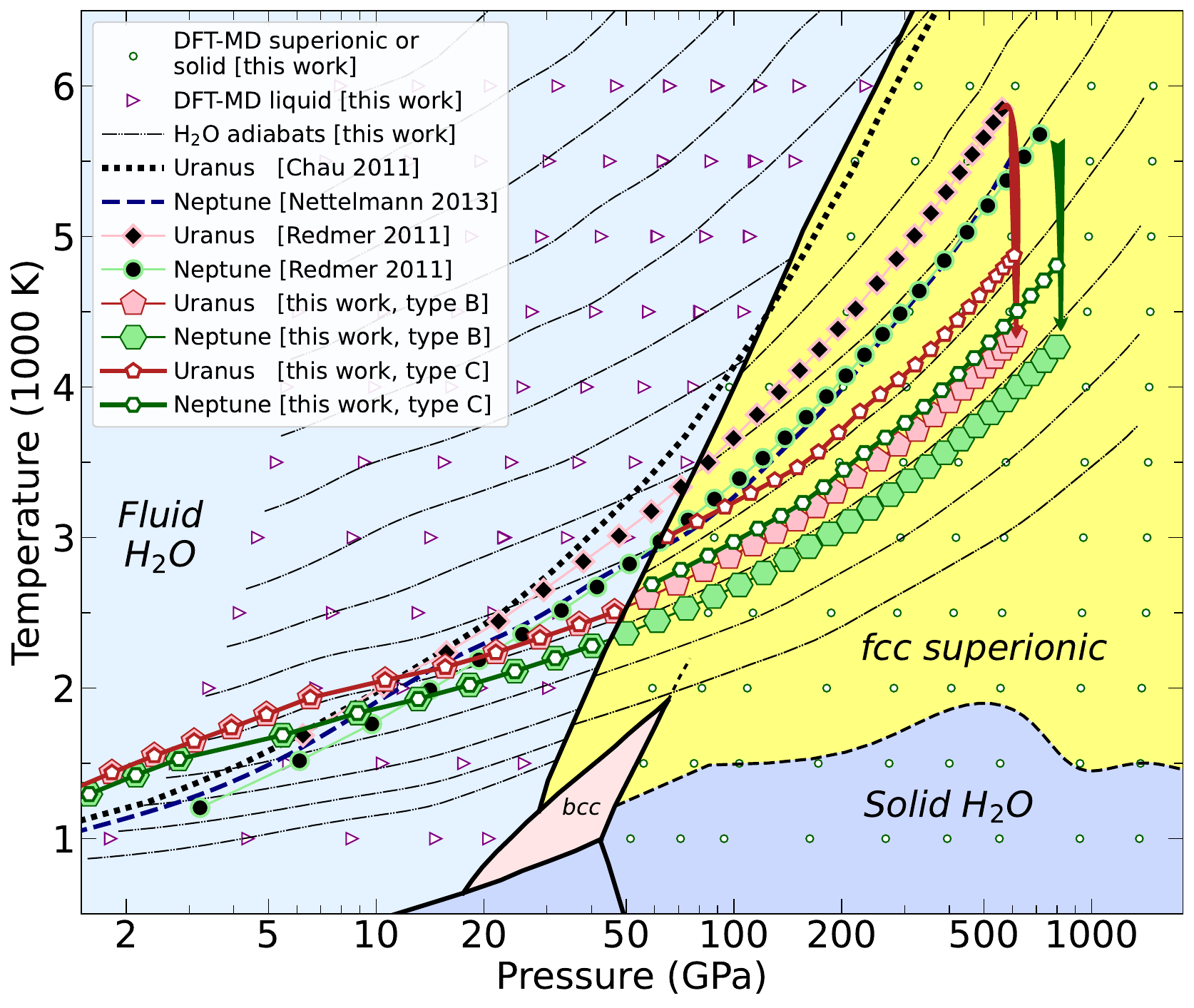}
\caption{Pressure-temperature profiles from various models for the interiors of Uranus and Neptune are plotted over the phase diagram of H$_2$O that was constructed by combining results from \citet{prakapenka2021structure} and \citet{WilsonWongMilitzer2013}. The slope of the adiabats (dash-dotted lines) that we derived with {\it ab initio} free energy calculations (triangles and circles) are much shallower than previous predictions by \citet{Chau2011} and \citet{Nettelmann2013}. The arrows emphasize that the temperatures of our continuous adiabats (models of type B) at the core-mantle boundaries are $\sim$1400~K lower than predicted by \citet{Redmer2011}. The deviations would have been even larger if \citet{Redmer2011} had not begun their adiabats at a 400~K lower temperature at 3~GPa. For this pressure, our models predict higher temperatures of 1630 and 1540~K for Uranus and Neptune, which is a result of differing adiabats in the hydrogen-helium layer in Fig.~\ref{fig:HHe}. Our models of type C assume that the liquid-to-superionic transition introduces a step into adiabats because of latent heat release.\label{fig:H2O}}
\end{figure}

The relevance of superionic water to planetary science was first
demonstrated by \citet{cavazzoni} who showed with {\it ab initio}
computer simulations that at high pressure and elevated temperatures,
water assumes a hydrid state, in which the larger oxygen atoms remain
confined to their lattice sites like atoms in a solid while the
smaller hydrogen nuclei diffuse through the oxygen sublattice like the
atoms of a liquid. \citet{french-prb-09} and \citet{Redmer2011}
performed simulations of a wider range of pressure-temperature
conditions and predicted superionic water to exist in the interiors of
Uranus and Neptune. At that point, all simulations of superionic water
had assumed a body-centered cubic (bcc) lattice at oxygen atoms
because \citet{cavazzoni} had initialized their simulations with solid
ice X structure. \citet{WilsonWongMilitzer2013} demonstrated with {\it
  ab initio} Gibbs free energy calculations that a more densely packed
face-centered cubic (fcc) structure is thermodynamically preferred for
most pressure-temperature conditions. With novel shock compression
experiments, \citet{millot2019nanosecond} confirmed the existence of
the superionic fcc phase.  Later diamond anvil cell experiments by
\citet{prakapenka2021structure} showed that the bcc structure exists
but it is confined to a rather small range of pressure-temperature
conditions and probably not relevant for ice giant interiors as we
illustrate in Fig.~\ref{fig:H2O}. Superionic behavior has also been
predicted in H$_2$O-NH$_3$ mixtures \citep{bethkenhagen2015superionic}
and will likely be a very common phenomenon among OCNH compounds
\citep{deVilla2023,deVilla2025}. At $\sim$20 Mbar, superionic fcc
structure has been predicted to transform into a different superionic
structure with $P2_1/c$ symmetry \citep{sun2015phase,militzer2018ab}
but the required pressures exceed the conditions in the interiors of
Uranus and Neptune.

In this paper, we report results from {\it ab initio} computer
simulations of liquid and fcc superionic water under conditions of ice
giant interiors. % Employing a thermodynamic integration technique enables us to compute absolute entropies for a given density and temperature. 
As we explain in the following Methods section, we first describe how
we calculate adiabats of mixtures of para and ortho hydrogen with
helium in the atmospheres of Uranus and Neptune, which enables us to
derive adiabats for their atmospheres. Then we describe our {\it ab
  initio} simulations and thermodynamic integration (TDI) method to
derive the entropy as a function of density and temperature, which
allows us to construct adiabatic temperature profiles for their icy
mantles.  Finally we explain how we construct models for the interiors
of Uranus and Neptune with the Concentric MacLaurin Spheroid (CMS)
method. We follow \citet{militzer2024phase} when we incorporate
results from atomistic {\it ab initio} simulations into planet-scale
CMS calculations to propose models for the interior structures of
Uranus and Neptune that are consistent with existing gravity
measurements and compatible with the proposed convective regimes.

In the Results section, we report our findings from of all three
methods. Our calculations predict the outer hydrogen-helium layers of
Uranus and Neptune to be hotter than earlier models
assumed. Conversely, our adiabats of liquid and superionic water much
shallower in pressure-temperature space than earlier predictions,
which means that we predict the deep interiors of Uranus and Neptune
to be 30\% colder than earlier models. If we change the model
assumptions and assume there is an offset in temperature between
liquid and superionic adiabats because of latent heat release, the
deviation from earlier models reduces to 15\%. Finally, we conclude by
discussing the implications of the proposed colder interiors for
future research on ice giant planets.

\section{Methods} \label{sec:methods}

\subsection{Para and Ortho Hydrogen}

In the outer atmospheres of giant planets, hydrogen and helium behave
like ideal gases. While interaction effects can be neglected, internal
degrees of freedom of the hydrogen molecules matter, which renders
the difference between para and ortho hydrogen important. The para
state is the ground state. It is not degenerate, its nuclear
spin wave function is antisymmetric ($S=0$), and its rotational wave
function is symmetric, which implies that only states with even
rotational quantum number $J$ can be occupied. Conversely, ortho
hydrogen has a symmetric nuclear spin wave function ($S=1$) with
degeneracy $2S+1=3$. Its rotational wave function is antisymmetric and
only states with odd $J$ can occupied. The para-to-ortho ratio is
temperature dependent, as the corresponding eigenstates have
different energies, but a temperature of 166.1~K in Jupiter's
atmosphere is already high enough for the para-to-ortho ratio to approach its
high-temperature limit of 1:3. The conversion between para and ortho
states is slow because collisions between molecules typically do not
flip nuclear spins. So when measurements in giant planet atmospheres
found the para-to-ortho ratios to deviate from the equilibrium value,
they provide an estimate for the speed of convection that brought up
gases from deep and hotter regions.

Overall one may assume the atmosphere of a giant planet be in thermal
equilibrium. To characterize its state, we follow the work by
\citet{saumon1991fluid}. Molecular hydrogen has 46 bound electronic
states, $\alpha$. Each has many vibrational states, $m$, and
rotational states, $J$ \citep{huber2013molecular}. Their energies have
been carefully characterized \citep{NIST_H2}:
\begin{eqnarray}
\label{eq:energyOfStates}
\epsilon(\alpha,m,J) = T_e 
+ \omega_e (m+\tfrac{1}{2}) 
- \omega_e x_e \left(m+\tfrac{1}{2}\right)^2 
+ B_e J(J+1) 
- D_e J^2(J+1)^2 
- \alpha_e  (m+\tfrac{1}{2}) J(J+1),
\,\,\, \,\,\, \,\,\,
\end{eqnarray}
where $T_e$ is the electronic groundstate energy, $\omega_e$ the first
vibrational constant, $\omega_e x_e$ the second vibrational constant,
$B_e$ the rotational constant, $D_e$ the centrifugal distortion
constant, and $\alpha_e$ is the rotational constant. They all depend
slightly on $\alpha$ \citep{NIST_H2}. In equilibrium one does not need
to treat para and ortho hydrogen as separate species but can instead
write down a single partition function that includes both nuclear spin
states $S=0$ and 1 with their corresponding degeneracy factors of
$g=1$ ($\Sigma_g$ and even $J$; $\Sigma_u$ and odd $J$), $g=3$
($\Sigma_g$ and odd $J$; $\Sigma_u$ and even $J$), and $g=4$
(electronic states $\Pi$ and $\Delta$):
\begin{eqnarray}
 Z = \sum_{\alpha,m,S,J} g(\alpha,J) \, \, (2J+1) \, \omega_{\alpha m} \, e^{-\epsilon(\alpha,m,J) / k_BT} \quad.
\label{eq:sumOfStates}
\end{eqnarray}
where the factor $(2J+1)$ represents the multiplicities of rotational
states. The factor $\omega_{\alpha m}$ represents the occupation
probabilities for given electronic and vibrational states, $\alpha$
and $m$. The sum can be simplified because even $J$ only contribute to
para states with $S=0$ and odd $J$ only matter for ortho states with
$S=1$. From this partition function, we can derive the internal
energy, entropy, and compute the para and ortho fractions as a function
of temperature. 
%
%By including translational degrees of freedom to compute the equation of state of a hydrogen-helium mixtures in the atmospheres of giant planets.
%
We have made our source code for this calculation
available~\citep{Zenodo_HHe}, which includes the
translational degrees of freedom and allows one to compute the
equation of state of a hydrogen-helium mixtures in the atmospheres of
giant planets. Under those conditions, one finds that most states in
Eq.~\ref{eq:sumOfStates} have too high energies to be occupied and
this calculation reduces to a sum over a surprisingly small number of
states. Most importantly, Eq.~\ref{eq:sumOfStates} enables one to
compute an absolute entropy to anchor the adiabats for the outer
layers of giant planets under low density conditions where it is
impractical to perform {\it ab initio} simulations. The two following
sections discuss them in detail because they are the preferred tool to
characterize materials at high pressure.

\subsection{Ab initio simulations}

To characterize H$_2$O in the mantles of Uranus and Neptune, we
performed density functional molecular dynamics (MD) simulations with the Vienna {\it Ab Initio} Package (VASP) \citep{VASP}. We employed
the Perdew, Burke, and Ernzerhof functional~\citep{PBE} and used hard
pseudo-potentials with the projector augmented-wave (PAW)
method~\citep{Kresse1999}. The valence configurations for the atoms
were O([He]2s$^2$2p$^4$) and H(1s$^1$). Following
\citet{WilsonWongMilitzer2013}, all simulations employed 144
atoms. Liquid simulations used cubic cells of the appropriate
volumes. For the superionic simulations with face-centered cubic
oxygen sublattice, we constructed monoclinic but nearly cubic
supercells with parameters $a=b$, $c/a=1.095$,
$\alpha=\beta=90^\circ$, and
$\gamma=78.5^\circ$~\citep{militzer2016supercell}. We consistently
employed a 2$\times$2$\times$2 Monkhorst-Pack grid to sample the
Brillouin zone. The electronic wave functions were expanded in a
plane-wave basis with an energy cut-off of 900~eV. All molecular
dynamics simulations used a time step of 0.2~fs to accommodate the
motion of the light hydrogen nuclei.  The temperature of our NVT
ensembles was regulated by a Nos\'e-Hoover thermostat~\citep{Nose1984}.

\subsection{Thermodynamic integration}

The entropy is a measure of the total number of microstates in an
ensemble and is thus not directly accessible with MD and MC
simulations, which only construct a representative subset of microstates
and do not attempt to perform an integration over all states. On the
other hand, MD and MC method are very good in calculating free energy
differences between two similar ensembles via thermodynamic integration, which
enables one to indirectly derive the entropy, $S$, from $S=(E-F)/T$,
where $E$ and $F$ are the internal and Helmholtz free energies. Like
pressure, the internal energy can be derived from a single MD or MC
simulation \citep{AT87}. Gaining access to the free energy requires
one to construct a thermodynamic integration path \citep{Wijs1998}
from a state of known free energy, $F_{id}$. For liquids, we choose an
ideal gas and, for solids, we use an Einstein crystal, in which every
particle is tethered to a particular lattice site. The integration
should be stable and efficient, which is why we employ the following
two-step integration procedure
\citep{Morales2009,WilsonMilitzer2012,WilsonMilitzer2012b,Militzer2013,Wahl2013a,Gonzalez-Cataldo2014,Wahl2015}
to derive the Helmholtz free energy, $F$, of the DFT system for given
temperature, volume, and particle number,
\begin{eqnarray}
\label{eq:TDI}
F_{DFT} &=& (F_{DFT}-F_{PP}) + (F_{PP}-F_{id}) + F_{id}\,,\\
&=& \int_0^1 d\lambda \left<U_{DFT}-U_{PP}\right>_\lambda
+ \int_0^1 d\lambda \left<U_{PP}-U_{id}\right>_\lambda
+ F_{id} \,.
\end{eqnarray}
The angular brackets represent an average for a single $\lambda$ value
that was computed by sampling particle configurations according to a
hydrid potential energy function,
$U_\lambda(r)=U_a(r)+\lambda (U_b(r)-U_a(r))$. Because DFT simulations
are orders of magnitude more expensive than calculations with pair
potentials (PP), one wants the first integration step in
Eq.~\ref{eq:TDI} to be most efficient so that only a few integration
steps are needed and the average of $U_{DFT}-U_{PP}$ converges
quickly. For given temperature, volume, and composition, we perform a
preliminary DFT-MD simulation and fit a pair potential to the computed
DFT forces~\citep{forcematching}. In this process, three points need to be considered: 

(1) For solids, the pair potentials and the Einstein potentials contribute to holding
particles in place while their combined forces should match the DFT
forces. So we choose pair and Einstein potentials such that each
contribute approximately half. For every atom type, we derive a
harmonic force constant, $k$, from
$3/2 k_B T = k/2 \left(\rr_i(t)-\rr_i^0\right)^2$ where $\rr_i(t)$
marks the position of particle $i$ and $\rr_i^0$ is its position in a
perfect lattice. We then lower $k$ by 50\%, subtract Einstein forces
from the DFT forces, and fit a pair potential to the remaining
forces. For liquids, pair potentials are fitted to the unaltered DFT
forces.

\begin{figure}[ht!]
%\gridline{\fig{plot_pair_potentials_02.pdf}{0.37\textwidth}{}}
%\gridline{\fig{plot_V_KS-V_PP_05.pdf}{0.39\textwidth}{}}
%\gridline{\fig{plot_V_vs_lambda_10.pdf}{0.37\textwidth}{}}
\gridline{\fig{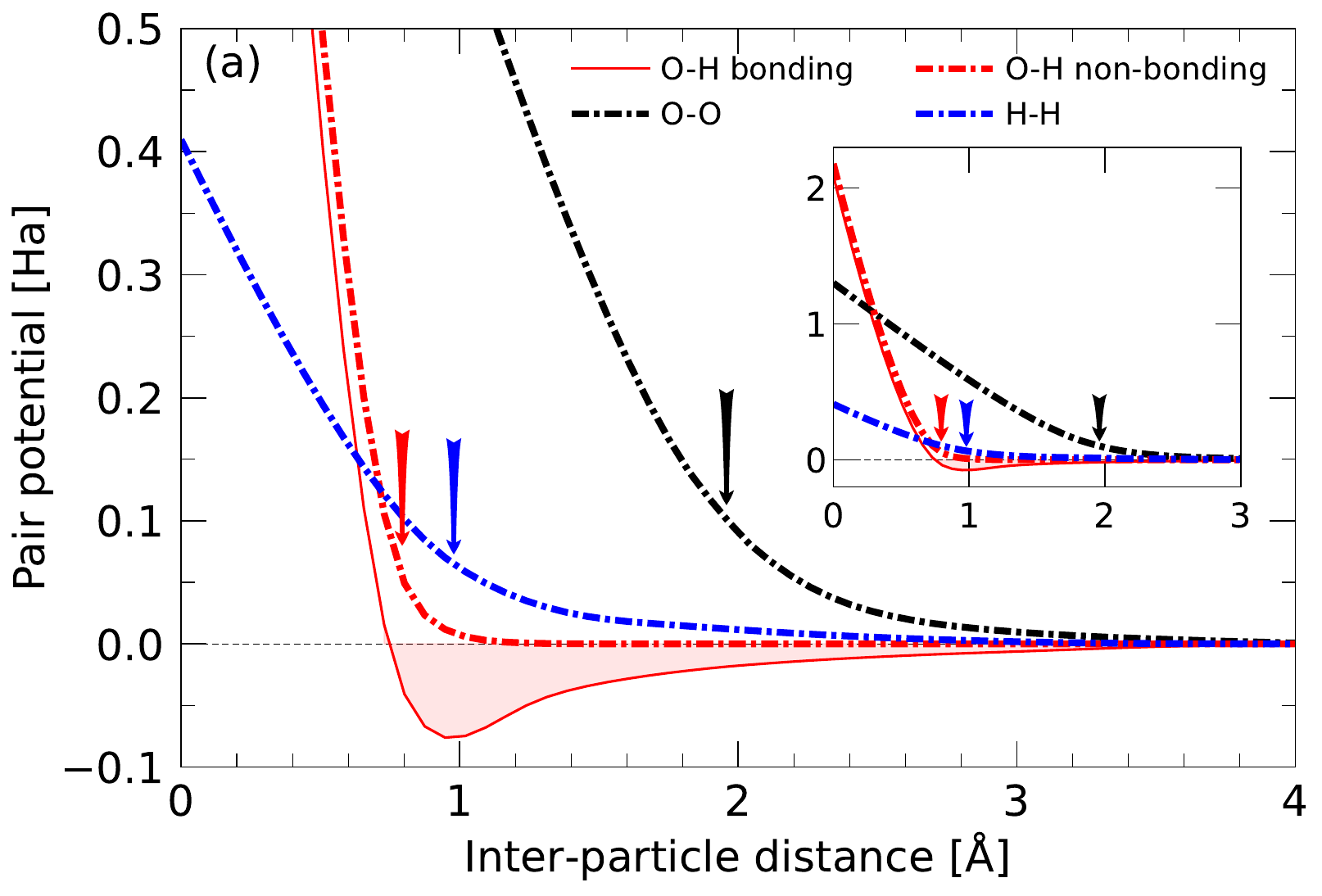}{0.38\textwidth}{}}
\vspace*{-9mm}
\gridline{\fig{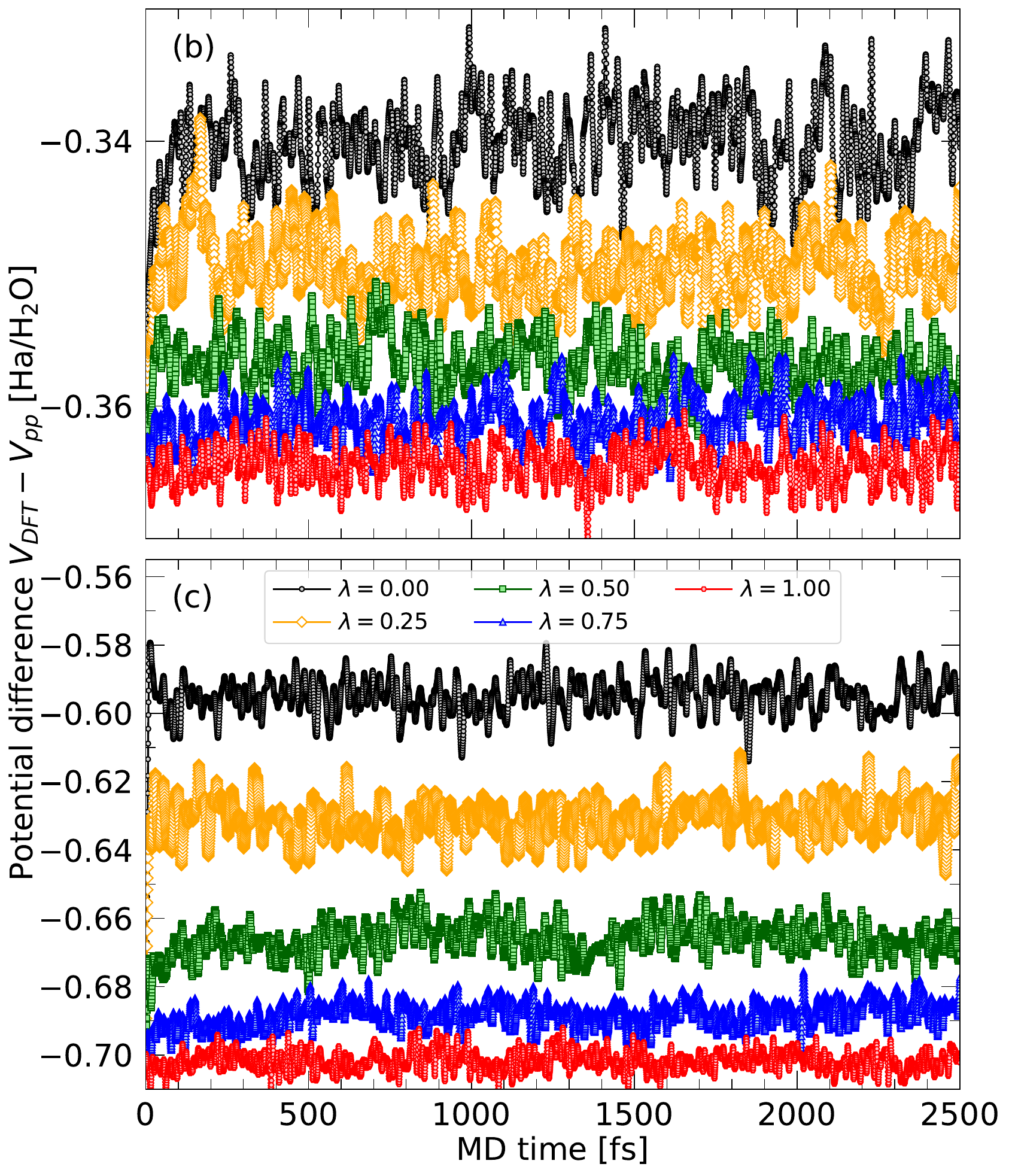}{0.40\textwidth}{}}
\vspace*{-9mm}
\gridline{\fig{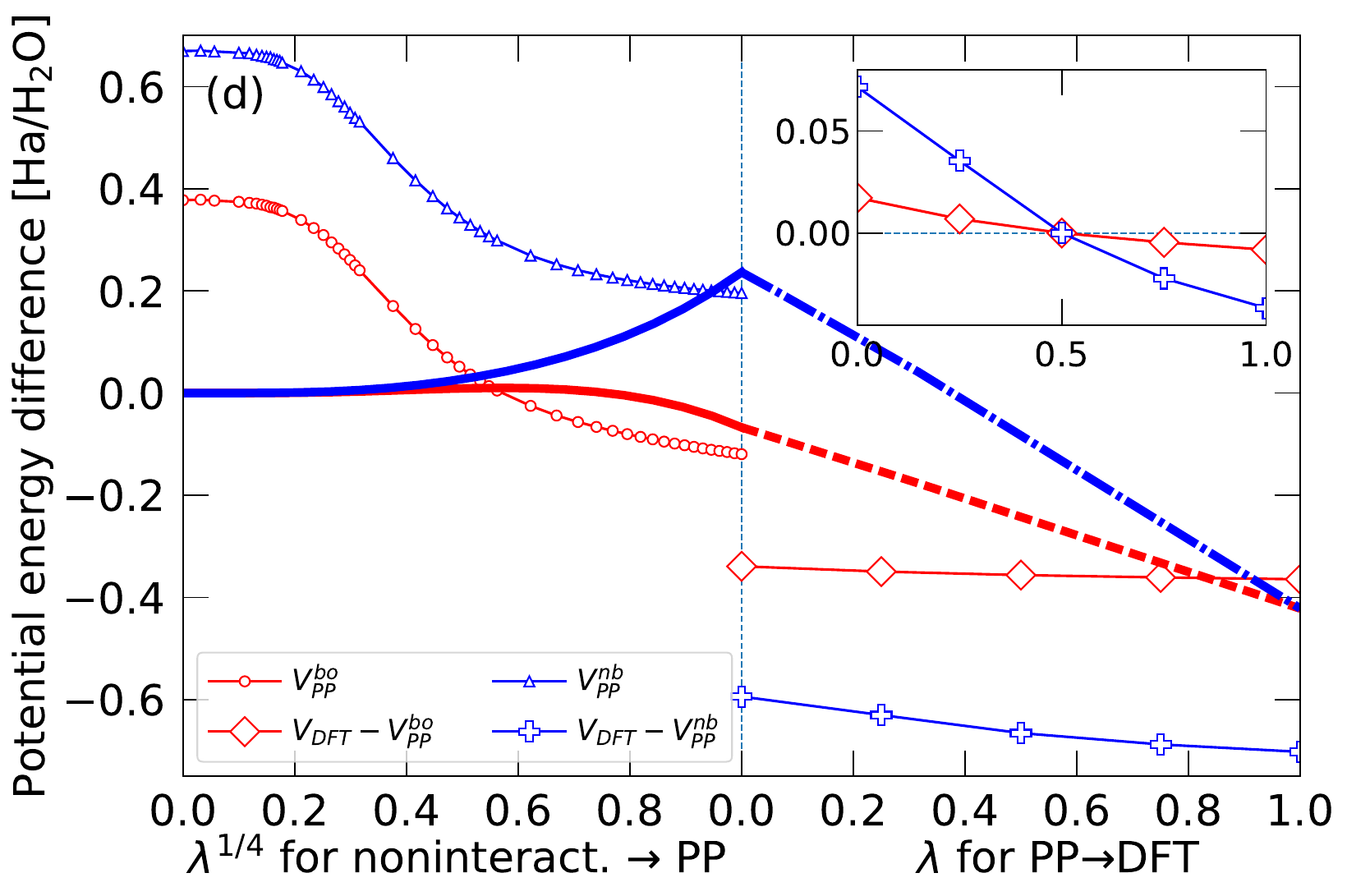}{0.38\textwidth}{}}
\vspace*{-9mm}
\caption{Panel (a) shows pair potentials between O and H atoms that were derived from a simulation of liquid H$_2$O at 3000~K, 1.75 g$\,$cm$^{-3}$, and 23~GPa. The arrows mark the beginning of the extrapolation towards smaller distances for each pair of atom types. For bonding and nonbonding pair potentials, panels (b) and (c) display the DFT and pair potential energy as function of MD time. Their time averages are represented by the diamond and plus symbols on the right side of panel (d). In the inset, we subtracted their respective $\lambda$=0.5 value from both functions to better compare changes in their magnitudes. The purpose of panel (d) is illustrate that one obtains the same integral when one performs the thermodynamic integration with bonding and nonbonding potentials. On the left side, we plot the average pair potential energies as function of $\lambda^{1/4}$ that we derived classical MC calculations. Their integrals over $\lambda$ are represented by the two thick lines. On right side, dashed and dot-dashed lines continue the integration into the regime where we switch from pair potentials to DFT forces. At $\lambda=1$, both lines converge to $-$0.42~Ha/H$_2$O emphasizing that both integration methods yield consistent results.\label{fig:TDI}}
\end{figure}

(2) Atoms repel each other strongly at close range because of Pauli
exclusion effects and Coulomb repulsion. There are for example no
pairs of oxygen atoms in the simulation in Fig.~\ref{fig:TDI} that are
closer than 1.9~$\AA$. The fitted pair potentials thus need to be
extrapolated towards small distances. The extrapolated part of the
pair potentials are not invoked in the first integration step in
Eq.~\ref{eq:TDI} but they become important in the second where we
gradually turn off the pair potentials. For liquids, the limit
$\lambda \to 0$ is equivalent to the limit of infinite temperature
where particles are uniformly distributed and the average pair
potential energy and its $\lambda$ derivative are given by
%\begin{eqnarray}
\begin{equation}
\label{eq:UPP}
\left< U_{PP} \right>_{\lambda=0} = \frac{4 \pi}{V} \int \, dr \, r^2 \, U_{PP}(r)
\quad{\rm and }\quad
\left. \frac{ \left< U_{PP} \right>} {d \lambda} \right|_{\lambda=0} = \beta \left<U_{PP}\right>^2 -  \beta \left<U_{PP}^2\right>
\quad{\rm with} \quad \beta = 1/k_BT\,.
\end{equation}
%\end{eqnarray}
The integrals are finite because we construct our potentials so that
they go to zero for large distances. In the limit of high temperature,
the particles become arbitrarily close, for which typical DFT codes
cannot construct electronic orbitals. This is another reason for why
we first switch from DFT to pair potentials in
Eq.~\ref{eq:TDI}. During construction, we extrapolate out pair
potentials linearly to small distances (see Fig.~\ref{fig:TDI}) so that
the value of $\left< U_{PP} \right>_{\lambda=0}$ in Eq.~\ref{eq:UPP}
remains finite.

(3) Pair potentials cannot accurately represent the nonadditive
many-body forces in hot, dense liquids and solids. However, 
Eq.~\ref{eq:TDI} yields exact results as long as the ensembles
generated by the DFT and by PP potentials are not too different, which
can become an issue for strongly bonded, molecular fluids like
hydrogen~\citep{Militzer2013}. The pair potentials will accurately
represent the strong intramolecular H$_2$ bond but then
indiscriminately applies it to all pairs of atoms neglecting all
intermolecular repulsion, which leads to the formation of unphysical
clusters because no many-body bonding effects are considered. To
prevent such instabilities, we follow \citet{Soubiran2015} and
remove the attractive part from the pair potentials by setting it
constant beyond the first potential minimum. Then we uniformly shift
the potential up so that approaches zero for large distances, as we show in
Fig.~\ref{fig:TDI}.

\subsection{CMS Calculations and Planetary Interior Models}

Following \citet{Nettelmann2013}, we adopted for Uranus and Neptune the equatorial radii of 25559 and 24766~km, planet masses of 14.536 and 17.148 Earth masses, rotation periods of 17:14:40 and 16:06:40 hours and 1~bar temperatures of 76 and 72~K. %The mass and equatorial radius also define a set of planetary units (PU) for each planet. 
The target values for gravitational moments were $J_2 \times 10^6 =  3510.99 \pm 0.72$ and $3529 \pm 45$ as well as $J_4 \times 10^6 =  -33.61 \pm 1$ and $-35.8 \pm 2.9$ as determined by the Voyager 2 spacecraft. We calculate these moments with,
\begin{equation}
  J_n = - \frac{4 \pi}{M a^n} \int\limits_{0}^{1} d \mu \int\limits_0^{r_{\rm max}(\mu)} \!\!\!\! dr \,\, r^{n+2} \,\, P_n(\mu) \,\, \rho(r,\mu) \quad,
  \label{standard_J}
\end{equation}
from the planet's interior density that we represent a function of radius, $r$,
and $\mu=\cos(\theta)$, the cosine of colatitude. $P_n$ are Legendre
polynomials, M is the planet's mass and $a$ its equatorial radius.

We construct models of the interior structure of Uranus and Neptune
with the accelerated version \citep{MilitzerSaturn2019} of the
nonperturbative Concentric MacLaurin Spheroid (CMS)
method~\citep{CMS}, which represents the interior of these rotating
planets by a series of $N_S=512$ axisymmetric spheroids. Their shapes
are adjusted until a state of hydrostatic equilibrium is established
that takes into account gravitational and centrifugal forces. The
rotation period and the equatorial radius are reproduced by
construction but matching the planet's total mass requires some
discussion. \citet{HubbardMilitzer2016} adjusted the density of the
innermost spheroid to match Jupiter's total mass. \citet{DiluteCore}
and \citet{Militzer_23456} varied the heavy element abundances of the
outer and inner layers to match Jupiter's mass and $J_2$. To match the
total mass of Uranus and Neptune, we scale the equatorial radii of the
three inner layers, $r_1$, $r_2$, and $r_3$ as the CMS method converges
to a hydrostatic solution. 

\begin{figure}[ht!]
\gridline{
\fig{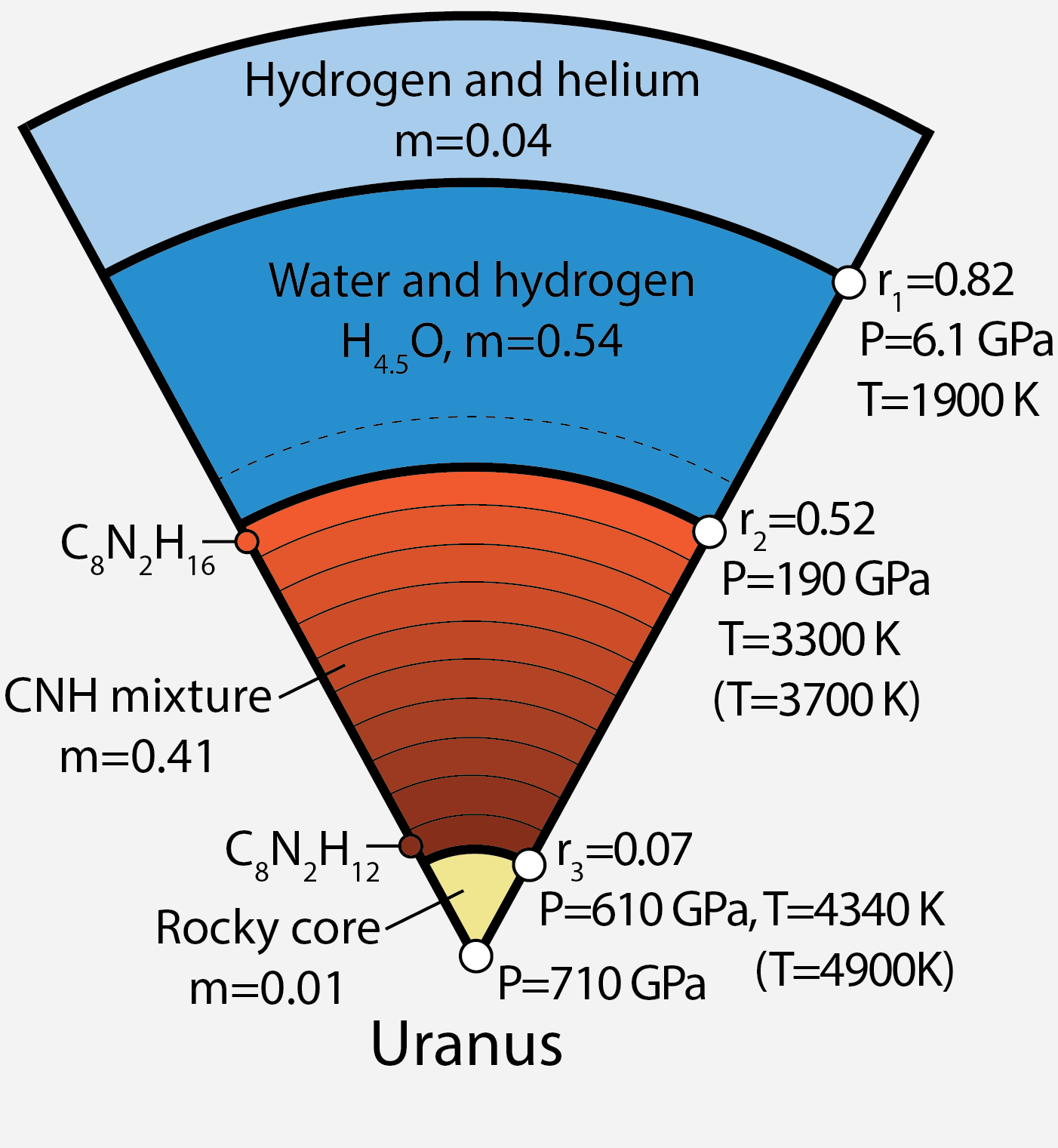}{0.37\textwidth}{}
\fig{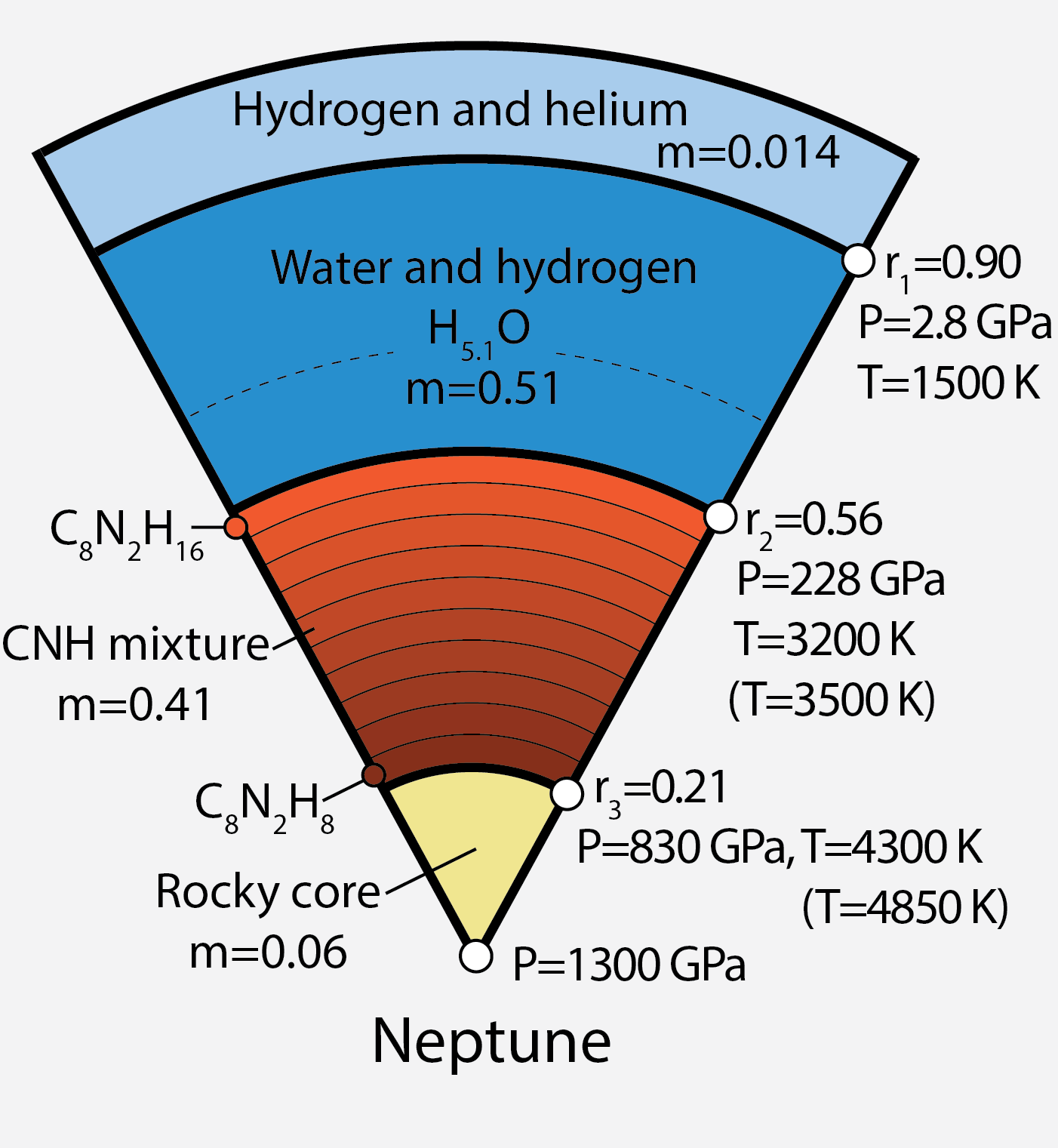}{0.37\textwidth}{}
}
\caption{Two representative interior models for Uranus and Neptune with four layers: 1) hydrogen and helium, 2) H$_2$O and hydrogen, 3) carbon-nitrogen-hydrogen mixture and 4) the rocky core composed of iron and rock. $m$ specifies the fractional layer masses. The temperatures for B and C type models are printed without and with parenthesis. The dashed lines in the second layer mark the conditions for the transition from liquid to superionic in pure H$_2$O.\label{fig:models}}
\end{figure}

As illustrated in Fig.~\ref{fig:models}, we followed
\citet{militzer2024phase} by constructing interior models with
four layers: 1) a protosolar mixture of hydrogen and helium, 2)
mixture of H$_2$O and hydrogen, 3) carbon-nitrogen-hydrogen mixture
and 4) the rocky core composed of iron and rock. We introduce the
parameters $H_1$, $H_2$, and $H_3$ to characterize the following
hydrogen fractions. We assume our ocean layer to be homogeneous and
convective, with $H_1 = N_H / 2 N_O$ defining its hydrogen fraction in terms
of the number of hydrogen and oxygen nuclei, $N_H$ and $N_O$. We
assume that the carbon-nitrogen-hydrogen layer is stably stratified
because it hydrogen contents vary between $H_2$ at the top and $H_3$
at the bottom. $H_{2,3} = N_H / (4 N_C + 3 N_N)$ are defined in
terms of the number of carbon and nitrogen atoms in the mixture, $N_C$
and $N_N$.

To reduce the discretization error, we slightly adjust the grid of the
equatorial spheroid radii so that every layer boundary coincides with
a grid point \citep{militzer2024phase}. We compared results of CMS
calculations with 256 and 512 spheroids and found that the change in
the predicted gravity coefficients is small. (\citet{Bailey_2021}
employed only $N_S=30$ spheroids for their Uranus and Neptune models.)
Scaling the radii $r_1$, $r_2$, and $r_3$ enables us derive a valid
interior model of the expected mass even in situations where one of
these three layers is rather small. This also means that we have
removed one dimension from our Markov chain Monte Carlo (MCMC)
calculations \citep{Militzer_QMC_2023,militzer2025ensemble} and now
have five independent parameters, $r_2/r_1$, $r_3/r_1$, $H_1$, $H_2$,
and $H_3$ that are constrained to satisfy
$1 \leq r_2/r_1 \leq r_3/r_1 \leq 0$ and $H_2>H_3$. The MC algorithm
is designed to generate ensembles of models by sampling from the
probability density, $\exp(-\chi^2/2)$, where $\chi^2$ represents the
deviation in the gravity coefficients between the observations and
model predictions that were computed with Eq.~\ref{standard_J},
\begin{equation}
\chi^2 = \sum_{n=1}^{2} \left( \frac{J_{2n}^{\rm model}-J_{2n}^{\rm obs}} {\delta J_{2n}^{\rm obs}} \right)^2
\quad.
\end{equation}
$\delta J_{2n}^{\rm obs}$ are one sigma error bars. 

Step by step the CMS algorithm \citep{CMS} converges to a hydrostatic
interior structure by repeatedly looking up information in the {\it ab
  initio} EOS tables. For example, the spheroid shapes are employed to
compute the potential on all spheroid surfaces that represents
gravitational and centrifugal forces. From these potential values, one
derives updated values for the pressure on all spheroids. For a given
layer and composition, one derives the temperature that correspond to
a pressure by assuming an isentrope of known entropy. Then one looks
up the corresponding density in the EOS table for given pressure,
temperature, and composition. The updated density profile is then
inserted into the CMS calculation to revise all surface potentials,
which enables one to update the spheroid shapes and pressure
values. We typically update the pressure values more frequently than
the spheoriod shapes because it allows us to converge to a
self-consistent hydrostatic structure more efficiently.

\section{Results} \label{sec:results}

\subsection{Hydrogen-helium layer}

\startlongtable
\begin{deluxetable}{c rrcccccc}
  \tablecaption{ Relationship of pressure, temperature and density for isentropes of the three giant planets that were derived with Eq.~\ref{eq:sumOfStates} for hydrogen-helium mixture with plausible helium mass fractions, $Y$, that are stated in column 1. Column 5 lists the entropy per particle, $N$ ($N=N_{\rm H_2} +N_{{\rm He}}$ for hydrogen molecules and helium atoms). In column 6, we list the related entropy per electron, $S^-/{\rm el}=[S/N]/2 - \Delta S \times N_{\rm H_2}/N $. The last three columns provide the fraction of para hydrogen and the fractional occupation of the first and second vibrational energy levels, $m=0$ and 1. \label{tab1}}
  \tablehead{ \colhead{ Planet } & \colhead{ P [bar] } & \colhead{ T [K] } & \colhead{$\rho$ [g$\,$cm$^{-3}$] } &
              \colhead{ S [k$_B$/N] } & 
              \colhead{ S$^-$ [k$_B$/el] } & \colhead{ f$_{\rm para\, H }$ } & 
              \colhead{ occ m=0 } & \colhead{ occ m=1 }  }
%\colnumbers
\startdata
Neptune  &     1 &   72.00 & 0.000391 & 12.635 & 5.737 & 0.5434 & 1.000 & 0.000 \\
Y=0.2777 &    10 &  145.13 & 0.001937 & 12.635 & 5.737 & 0.2910 & 1.000 & 0.000 \\
         &   100 &  298.71 & 0.009414 & 12.635 & 5.737 & 0.2507 & 1.000 & 0.000 \\
         &  1000 &  594.50 & 0.047299 & 12.635 & 5.737 & 0.2500 & 1.000 & 0.000 \\
         & 10000 & 1166.54 & 0.241047 & 12.635 & 5.737 & 0.2500 & 0.994 & 0.006 \\
\hline										  
Uranus   &     1 &   76.00 & 0.000370 & 12.834 & 5.836 & 0.5128 & 1.000 & 0.000 \\
Y=0.2777 &    10 &  154.82 & 0.001816 & 12.834 & 5.836 & 0.2818 & 1.000 & 0.000 \\
         &   100 &  317.18 & 0.008865 & 12.834 & 5.836 & 0.2505 & 1.000 & 0.000 \\
         &  1000 &  630.75 & 0.044581 & 12.834 & 5.836 & 0.2500 & 1.000 & 0.000 \\
         & 10000 & 1234.08 & 0.227855 & 12.834 & 5.836 & 0.2500 & 0.992 & 0.008 \\
%\hline										  
%Saturn   &     1 &  142.65 & 0.000193 & 14.874 & 6.838 & 0.2937 & 1.000 & 0.000 \\
%Y=0.238  &    10 &  292.58 & 0.000940 & 14.874 & 6.838 & 0.2509 & 1.000 & 0.000 \\
%         &   100 &  579.36 & 0.004745 & 14.874 & 6.838 & 0.2500 & 1.000 & 0.000 \\
%         &  1000 & 1131.95 & 0.024286 & 14.874 & 6.838 & 0.2500 & 0.995 & 0.005 \\
%         & 10000 & 2094.88 & 0.131228 & 14.874 & 6.838 & 0.2500 & 0.938 & 0.057 \\
\hline
Jupiter  &     1 &  166.10 & 0.000166 & 15.346 & 7.074 & 0.2736 & 1.000 & 0.000 \\
Y=0.238  &    10 &  336.96 & 0.000816 & 15.346 & 7.074 & 0.2503 & 1.000 & 0.000 \\
         &   100 &  665.98 & 0.004128 & 15.346 & 7.074 & 0.2500 & 1.000 & 0.000 \\
         &  1000 & 1291.94 & 0.021278 & 15.346 & 7.074 & 0.2500 & 0.990 & 0.010 \\
         & 10000 & 2355.50 & 0.116708 & 15.346 & 7.074 & 0.2500 & 0.915 & 0.077 \\
\enddata
\end{deluxetable}

\begin{figure}[ht!]
\plotone{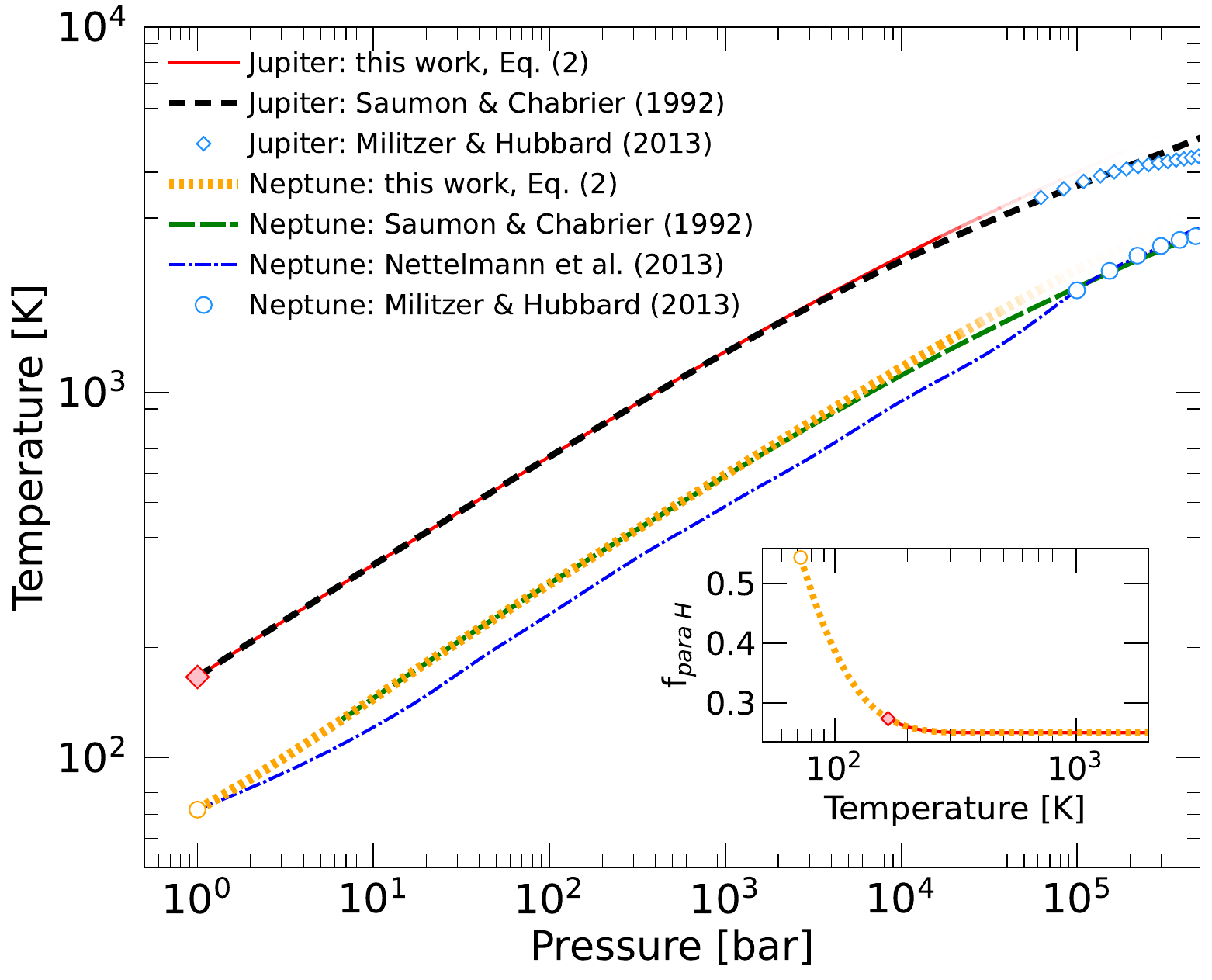}
\caption{Isentropes of hydrogen-helium mixtures in the outer layers of
  Jupiter and Neptune. The symbols at 1 bar mark the anchor points
  with temperatures of 166.1 and 72~K respectively. The inset shows
  the fraction of para hydrogen, which approaches 1/4 in the limits of
  high temperature. \label{fig:HHe}}
\end{figure}

We begin by discussing the temperature profile of Uranus and Neptune's
outer hydrogen-helium layer. We assume an isentrope with a protosolar
helium fraction of Y=0.2777 and neglect minor species such as
methane. In Fig.~\ref{fig:HHe}, we compare the temperature profiles
from Eq.~\ref{eq:sumOfStates} with other calculations. In
Tab.~\ref{tab1}, we state the anchor points at 1 bar for the
isentropes of different planets. For all higher pressures, the density
and temperature were adjusted until the target pressure and the
entropy of the anchor point were reproduced. What one might consider
to be a complex statistical problem of atomic physics greatly
simplifies under the conditions in the outer envelopes of giant
planets where the temperature is too low for many molecular
eigenstates to be occupied. For example, none of the excited
electronic states matter because already the first excited state
requires an excitation energy of 11.4~eV or 132$\,$000~K. Furthermore,
the last two columns of Tab.~\ref{tab1} show that only the lowest
vibrational energy is occupied unless the temperature reaches
$\sim$1000~K, at which the occupation fraction of the vibrational
$m=1$ state reaches about 0.3\%. This bring us to the rotational
states, which require the least excitation energy of $\sim$170~K. Even
for those, the corresponding sum in Eq.~\ref{eq:sumOfStates} converges
rapidly. For a temperature of 1000~K, just 12 rotational states are
required to compute the entropy with an accuracy of 10$^{-4}$
k$_B$/particle. The inset of Fig.~\ref{fig:HHe} shows the fraction of
para hydrogen, which will approach 1 in the limit of low
temperature. By 300~K, the para-to-ortho ratio has already converged
its high-temperature limit of 1:3 ($f_{\rm para\,H}=1/4$). At 1~bar,
the para fraction of Uranus and Neptune's atmosphere is about 50\%,
while it is about 27\% for Jupiter because its atmosphere is hotter.

In Fig.~\ref{fig:HHe}, we compare the isentrope predicted by
Eq.~\ref{eq:sumOfStates} with results in the literature. For pressures
up to 5~kbar, we find excellent agreement with the work by
\citet{SC92} (SC), for which a code is available that interpolates
precomputed tables for temperatures greater than 125~K. To make such
calculations more accessible and to extend them to lower temperatures,
we make our source code available~\citep{Zenodo_HHe}. For
pressures exceeding 5~kbar, the interaction between particles that we
have neglected in Eq.~\ref{eq:sumOfStates} start to
matter. \citet{SC92} included different types of interactions into
their EOS tables, which reduce the temperature of an isentrope. At a
pressure of 100~kbar (10~GPa), their predictions converged to the
results from {\it ab initio} simulations from \citet{MH13} (MH13) for
Neptune and Jupiter as Fig.~\ref{fig:HHe} shows.

For the comparison between {\it ab initio} simulations and analytical
methods like Eq.~\ref{eq:sumOfStates} to be meaningful, one needs to
agree on a definition for the entropy and both methods need to be able
to provide absolute entropies. The {\it ab initio} entropies of MH13
were derived with a TDI method that relied on classical nuclei and
quantum mechanical forces and therefore does not include any spin
effects of the hydrogen nuclei. This is perfectly reasonable at high
pressure where the nuclear spin states do not affect the motion of the
atoms. In the weakly coupled regime at low pressure, nuclei spin effects need to be
considered because they affect the rotational states of the hydrogen
molecules. The simplest way to make the {\it ab initio} entropies of
MH13 compatible with entropies of Eq.~\ref{eq:sumOfStates} and of SC
EOS is to add $\Delta S= k_B \ln(2)$ per hydrogen atom to the MH13
entropies to incorporate the missing nuclear spin degrees of
freedom. If one does not want to alter the MH13 entropies for some
reason one may subtract $\Delta S$ per atom from all entropies that
were computed with Eq.~\ref{eq:sumOfStates} (see $S^-$ in
Tab.~\ref{tab1}). This subtraction allows one to compare the entropy
values and to align the isentropes in P-T space but it would not yield
a proper thermodynamic entropy because it would approach $-\Delta S$
in the limit of low temperature.

Fig.~\ref{fig:HHe} shows that the SC predictions for Jupiter start to
deviate from the {\it ab initio} simulations above 200~kbar because
molecules start to interact strongly and one approaches the regime of
pressure driven molecular dissociation, which SC model interpolates
across. More details are provided in \citet{guillot-book,MH08,MHVTB}.

Fig.~\ref{fig:HHe} also shows that the Neptune model by
\citet{Nettelmann2013} predicts temperatures that are $\sim$20\% lower
than other predictions over a wide range of pressure from 10 bar to 20
kbar. % but then agrees with them very well from 100 to 500 kbar.

\subsection{TDI for liquid water}

In Fig.~\ref{fig:TDI}, we compare the results from two different TDI
calculations for liquid water at 3000~K, 1.75 g$\,$cm$^{-3}$, and
23~GPa. Panel (a) shows that only the O-H potential has an attractive
part, which we have removed when we constructed our set of nonbonding
potentials while the O-O and H-H were the same as in our set of
bonding potentials. Panels (b) and (c) demonstrate that TDI
simulations with bonding and nonbonding pair potentials are stable for
all $\lambda$ values. The ensemble generated with the nonbonding
potentials differs more from the DFT ensemble as the inset of panel
(d) illustrates. For the nonbonding potentials the integrand,
$\left<U_{DFT}-U_{PP}\right>_\lambda$, has more curvature and the difference
between $\lambda$=0 and 1 points is 0.11 Ha/cell while it is only
0.025 Ha/cell for our bonding potentials, which makes the time
consuming evaluation of the first term in Eq.~\ref{eq:TDI} more
efficient.

With the bonding potentials, we obtained an entropy of 21.97$\pm$0.26
k$_B$/H$_2$O while we derived 22.04$\pm$0.30 with the nonbonding
potentials. (An error bar of 0.3 k$_B$/H$_2$O translates into a small
temperature uncertainty of only 11~K.) Both results are consistent
with each other, which we further analyze in panel (d) of
Fig.~\ref{fig:TDI}. On the left side, we compare the two
$\left<V_{PP}\right>_\lambda$ curves (red and blue symbols). They
have similar but not identical shapes. The average potential energy,
$\left<V_{PP}\right>_\lambda$ is higher for the nonbonding
potentials because we shifted $U_{O-H}$ up when we removed the
attractive part of the potential. The integral
$\int_0^1 d\lambda \left<V_{PP}\right>_\lambda$ (solid blue line) is
thus larger than that of the bonding potential (red solid line). But when we
add the integral
$\int_0^1 d\lambda \left<V_{DFT}-V_{PP}\right>_\lambda$ in the right
figure panel, results with both pair potentials converge to the same
value of $-$0.42~Ha/H$_2$O at $\lambda=1$. This confirms that one
obtains consistent results for different classical potentials, which
is one of the main strength of the TDI method.

\subsection{Isentropes of liquid and superionic H$_2$O}
\begin{figure}[ht!]
\plotone{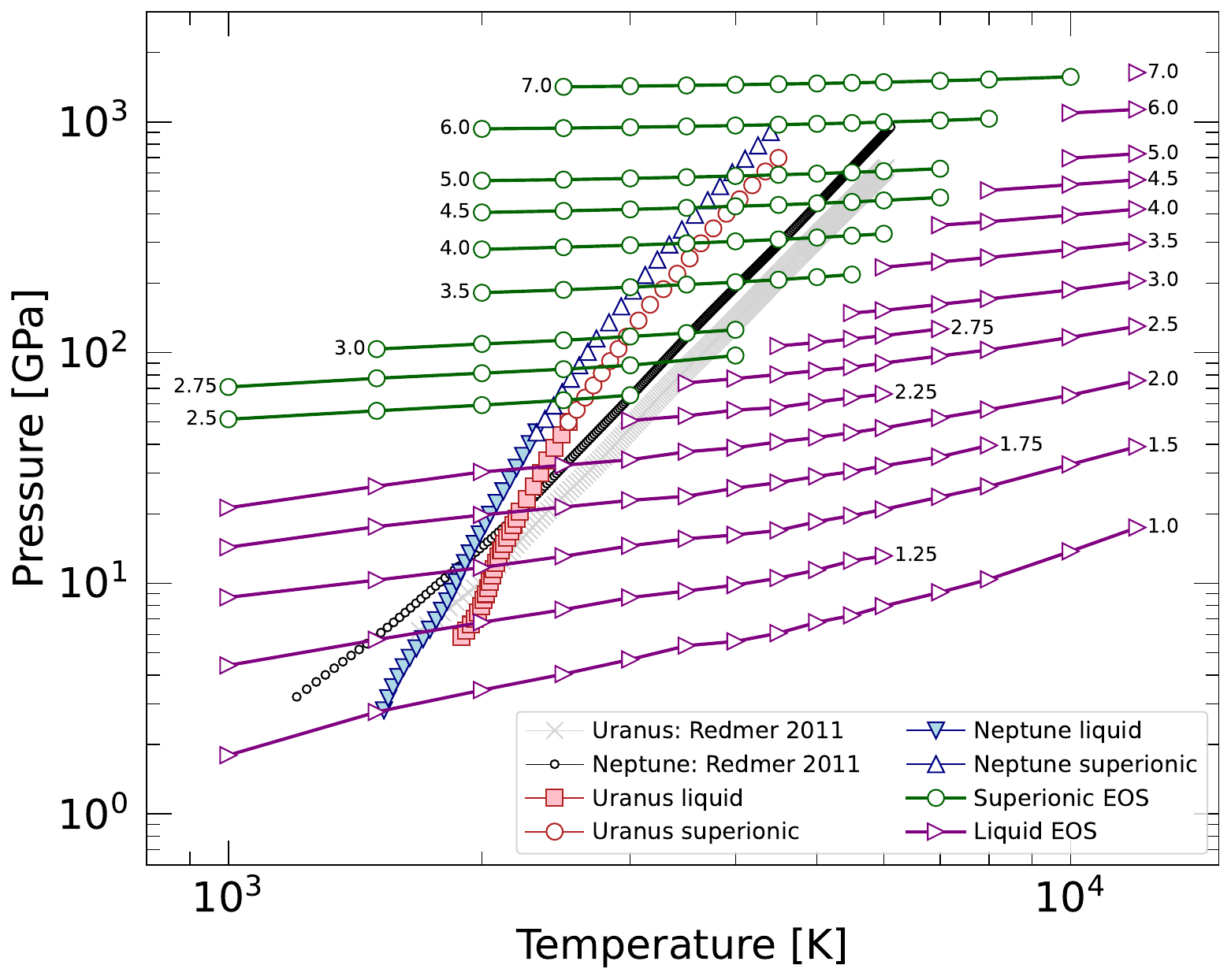}
\caption{Pressure for isochores of liquid and superionic H$_2$O. The labels specify the density in \gcc. The offset between the two families of curves implies a first order phase transition. For given density and temperature, the pressure of the liquid phase is higher than that of the superionic phase, which implies the melting line has a positive slope $dP/dT>0$. The temperature-pressure profiles of our interior models for Uranus and Neptune of type B are compared with the much hotter conditions proposed by \citet{Redmer2011}. \label{fig:T-P}}
\end{figure}

\begin{figure}[ht!]
\plotone{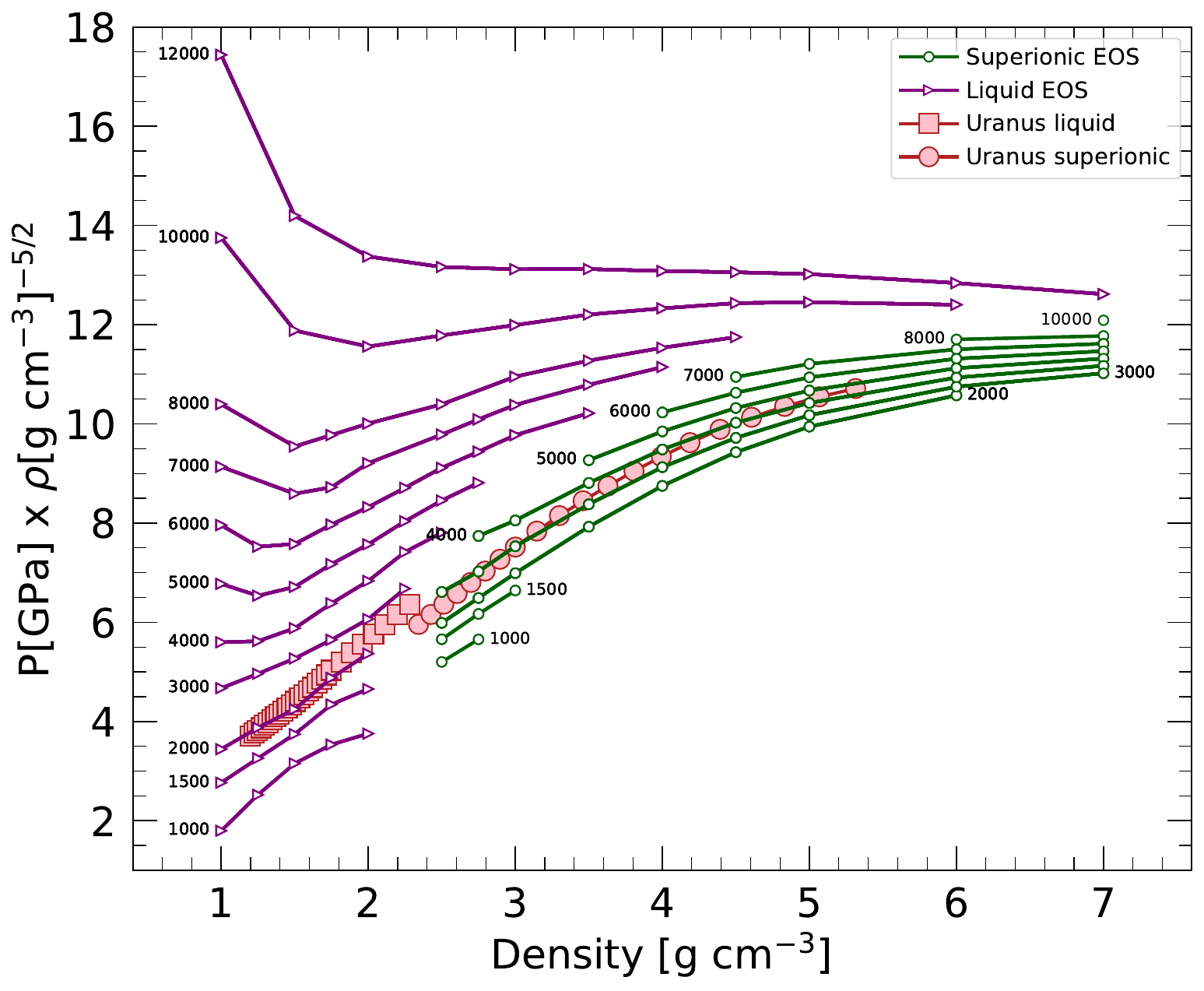}
\caption{Pressure of liquid and superionic H$_2$O is plotted as function of density for different temperature values that are specified on every curve in units of Kelvin. For clarity, we multiplied the pressure by density to the power of $-2/5$ because pressure varies strongly with density. Our Uranus model of type B from Fig.~\ref{fig:models} is included.\label{fig:rho-P}}
\end{figure}

\begin{figure}[ht!]
\plotone{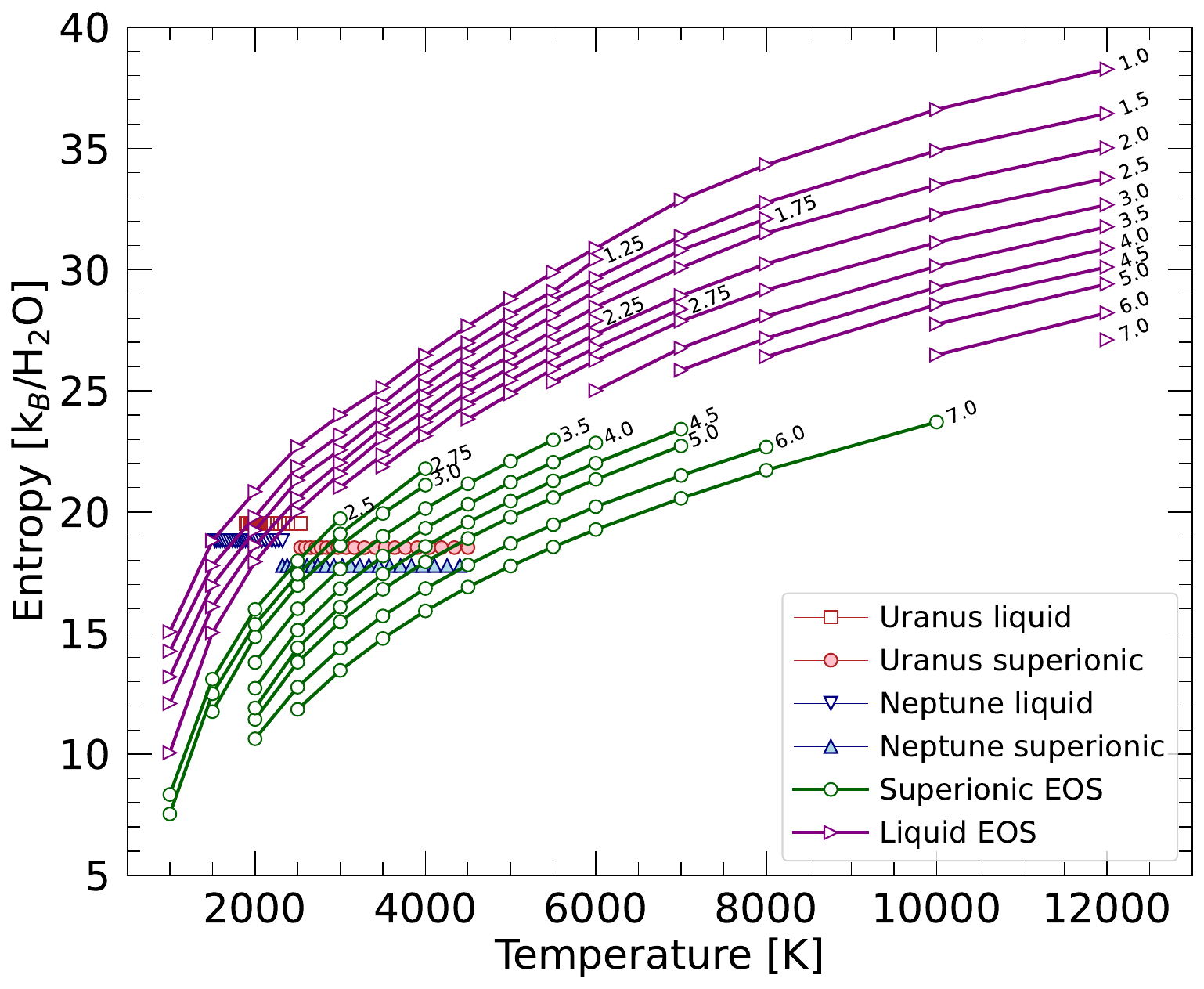}
\caption{Entropy of liquid and superionic H$_2$O that we computed for different density values that are specified in units of \gcc. For given density and temperature, the entropy of liquid water is higher because the oxygen nuclei are more disordered. For reference, we mark the conditions of our two interior models of type B for Uranus and Neptune in Fig.~\ref{fig:models}. \label{fig:T-S}}
\end{figure}

\begin{figure}[ht!]
\plotone{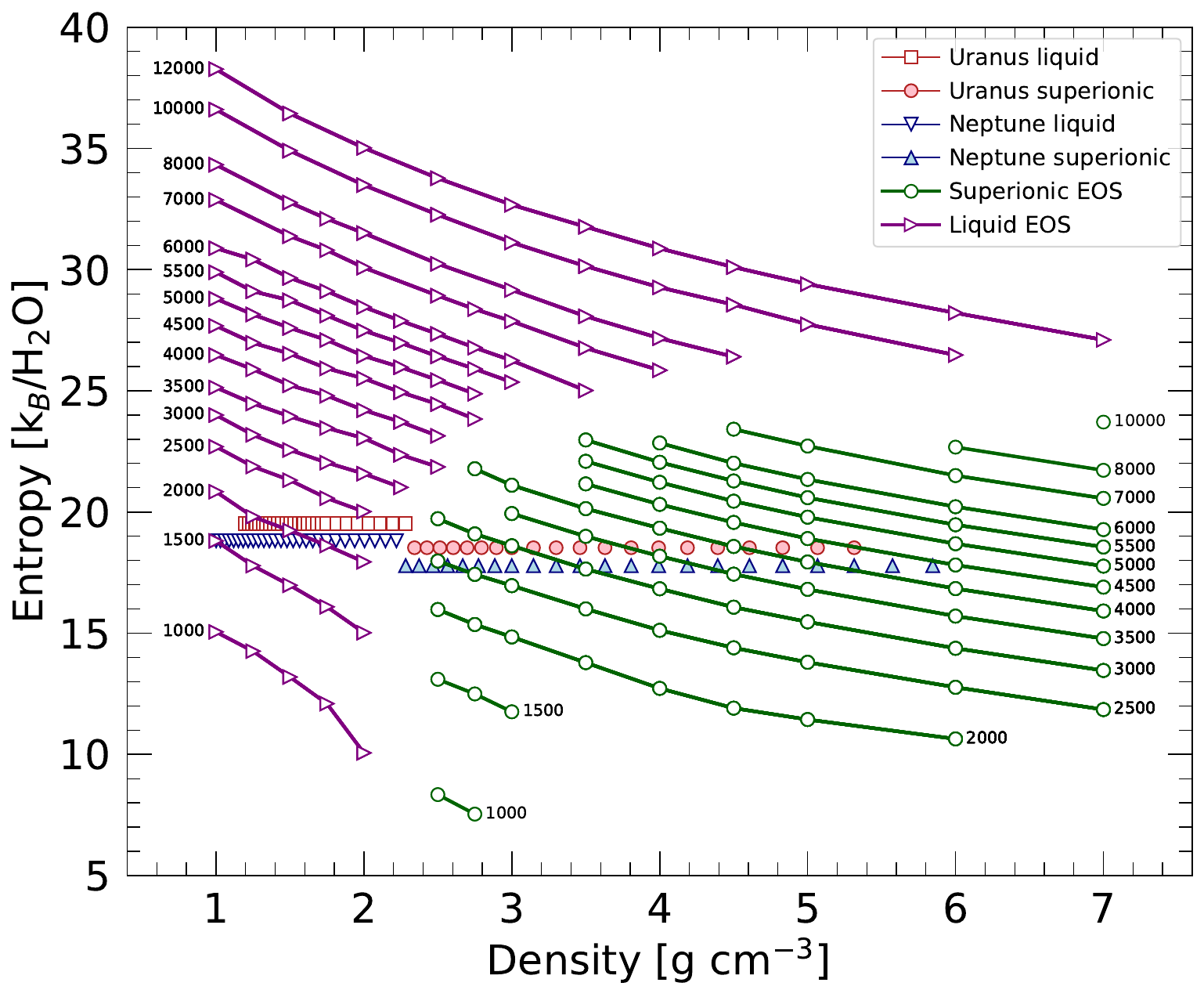}
\caption{Entropy of liquid and superionic H$_2$O from Fig.~\ref{fig:T-S} for different temperatures are plotted as function of density. For a given temperature, the liquid and superionic curves are offset because the oxygen nuclei are more disordered in the liquid phase. The conditions of our two interior models of type B for Uranus and Neptune in Fig.~\ref{fig:models} are included. \label{fig:rho-S}}
\end{figure}

%\begin{figure}[ht!]
%\plotone{plot_rho-S_07.pdf}
%\caption{xxxx \label{fig:T-P}}
%\end{figure}

In Figs.~\ref{fig:T-P} and \ref{fig:rho-P}, we plot the pressure that
we derived with our {\it ab initio} simulations over a wide range
density and temperature conditions that include the interiors of
Uranus, Neptune, and far beyond as we illustrate in
Fig.~\ref{fig:H2O}. Our equation of state table is available at
\citet{Zenodo_H2O}.

There is an offset in pressure between the liquid and superionic
branches as the 2.5 and 2.75 \gcc ~isochores in Fig.~\ref{fig:T-P}
illustrate. This implies the melting transition of superionic ice is a
first-order phase transition. It has a positive $dT/dP$ Clapeyron
slope because the superionic phase is denser. This is confirmed by the
offset in pressure between the 3000 and 4000~K isotherms in
Fig.~\ref{fig:rho-P}.

A first-order phase transition also implies a discontinuous change in
entropy as we illustrate in Figs.~\ref{fig:T-S} and \ref{fig:rho-S}
where we plot {\it ab initio} entropies that we derived for both
phases. Because oxygen atoms are more ordered in the superionic
structure, this phase has lower entropy, which can be seen best by
following the 2000, 3000, and 4000~K isotherms in
Fig.~\ref{fig:rho-P}.
So if a parcel of fluid H$_2$O transitions into the superionic state,
it would release latent heat and its density would decrease.

\subsection{Implications for Uranus and Neptune}

Figs. \ref{fig:H2O} and \ref{fig:T-P} show that our {\it ab initio}
derived isentropes are much shallower in pressure-temperature space
than those proposed earlier. At any point, their slopes can be
expressed in terms of
$\delta = \frac{\partial \ln T}{\partial \ln P}|_S$ and Gr\"uneisen
parameters $\gamma_G = \frac{\partial \ln T}{\partial \ln
  \rho}|_S$. If one considers a pressure or density range, one can
introduce effective values for these coefficients by fitting
$T \propto P^\delta$ and $T \propto \rho^{\gamma_G}$. For an ideal gas
of nonlinear, triatomic molecules, one obtains $\delta = 1/4$ and
$\gamma_G = 1/3$ by assuming the vibrational degrees of freedom are
frozen while the transitional and rotational motion can proceed
unhindered, which is unrealistic for fluids at high pressure where
interaction effects are very important.

For Uranus and Neptune conditions, we obtained $\delta \approx 0.15$
and $\gamma_G \approx 0.52$ for fluid H$_2$O from 3--20~GPa and
$\delta \approx 0.24$ and $\gamma_G \approx 0.73$ for superionic
H$_2$O from 130--1300~GPa. These values are much smaller than previous
predictions as the slopes of different curves in Fig.~\ref{fig:H2O}
illustrate. \citet{hubbard1980structure} assumed $\delta \approx 0.31$
and $\gamma_G \approx 0.9$ for the entire ice layer. \citet{Chau2011}
assumed a similar slope of $\delta \approx 0.32$. The isentropes by
\citet{Redmer2011} are consistent with a value of
$\delta \approx 0.28$. The adiabat of \citet{Nettelmann2013} can be
represented by $\delta \approx 0.23$ and $\gamma_G \approx 0.63$ from
10--60~GPa and by $\delta \approx 0.29$ and $\gamma_G \approx 0.82$
from 100--590~GPa.

To understand the likely implications of these shallower slopes for
the interiors for Uranus and Neptune we need to introduce additional
approximations because their mantles are in all likelihood not
composed of pure water \citep{Podolak1995}. So we adopted the interior
models from \citet{militzer2024phase} in Fig.~\ref{fig:models} and
employed our H$_2$O isentropes to approximately present the
temperature profiles of the outer water-hydrogen layer and of inner
carbon-nitrogen-hydrogen layer. Even though our {\it ab initio}
derived adiabats are colder than previously assumed, we had no
difficulties constructing models that match the observed gravity
fields of Uranus and Neptune. Two interior models are illustrated in
Figs.~\ref{fig:H2O} and \ref{fig:models}. The corresponding data files
are available at \citet{Zenodo_H2O}. 

\begin{figure}[ht!]
%\plotone{plot_Jn_01.png}
%\gridline{\fig{plot_Jn_01.png}{0.6\textwidth}{}}
%\gridline{\fig{plot_Jn_01_100dpi.png}{0.6\textwidth}{}}
%\gridline{\fig{plot_Jn_02_100dpi.png}{0.6\textwidth}{}}
\gridline{\fig{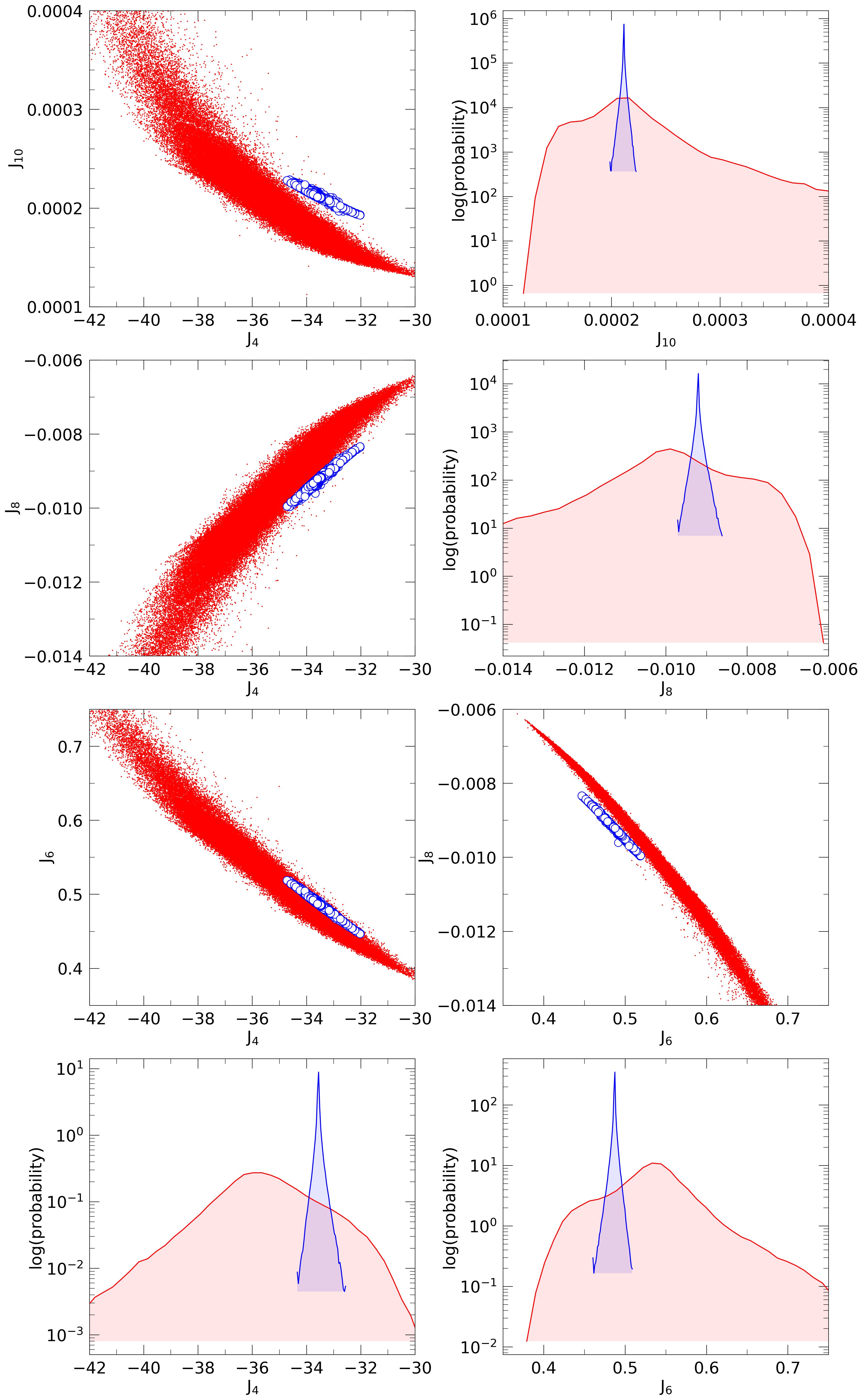}{0.6\textwidth}{}}
\caption{Posterior distributions of gravity harmonics, $J_n$, that we derived with MCMC calculations of the interiors of Uranus (blue) and Neptune (red). We show correlation plots and histograms. For the latter we chose a logarithmic Y axis to accommodate a wide range of probability densities. The panels were ordered so that some X axes align vertically.  No dynamic contributions from winds were considered in the calculation of the $J_n$. Predictions for Neptune vary much more because the planet's gravity field has been measured less accurately.\label{fig:Jn}}
\end{figure}

\begin{figure}[ht!]
% \plotone{correlation_plot_72k.png}
% \plotone{correlation_plot_72k_100dpi.png}
% \plotone{correlation_plot_72l_100dpi.png}
% \plotone{correlation_plot_73q_100dpi.png}
\plotone{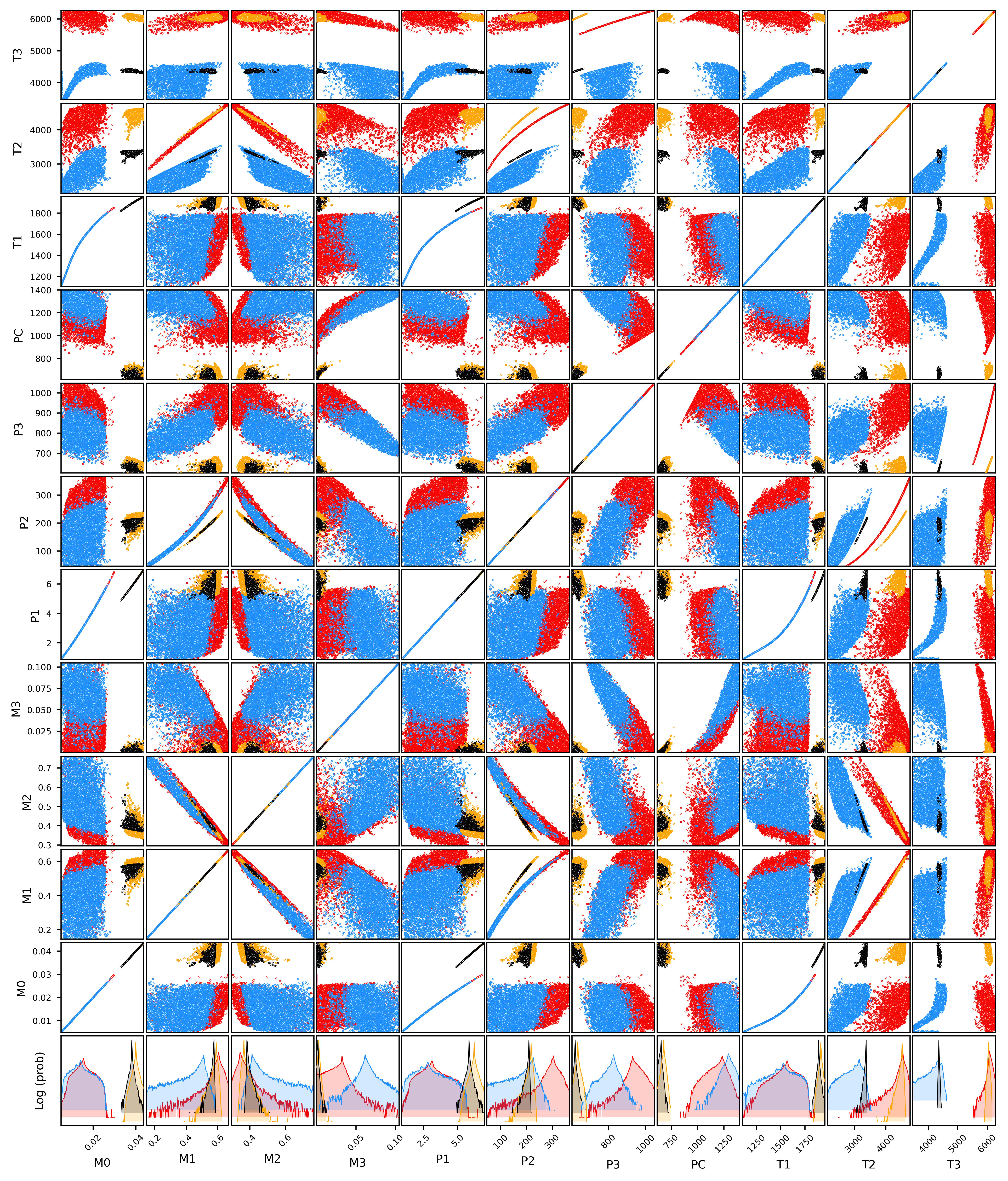}
% \plotone{correlation_plot_73q.png}
\caption{Posterior distributions of layer masses, boundary pressures and temperatures. M$_0$, M$_1$, M$_2$, and M$_3$ refer to the fractional masses of the four interior layers in Fig.~\ref{fig:models}: 1) H-He, 2) H$_2$O rich, 3) C-N-H and 4) the rocky core. $P_1$ in units of GPa and $T_1$ in K refer to the conditions of boundary between layers 1 and 2. Similarly, $P_{i>1}$ and $T_{i>1}$ characterize the boundaries between deeper layers in Fig.~\ref{fig:models}. Finally $P_C$ refers to the pressure at the planet's very center. The blue and red points respectively represent CMS ensembles of Neptune that we constructed with our temperature-pressure profile of type B and the hotter profile that was proposed by \citet{Redmer2011} that we show in Fig.~\ref{fig:H2O}. Conversely the black and orange symbols shows ensembles of Uranus models with our type B and \citet{Redmer2011} temperature profiles. Our temperatures are much lower than that by \citet{Redmer2011}, which affects many other planetary properties.\label{fig:corr}}
\end{figure}

\begin{figure}[ht!]
% \plotone{correlation_plot_74r_100dpi.png}
\plotone{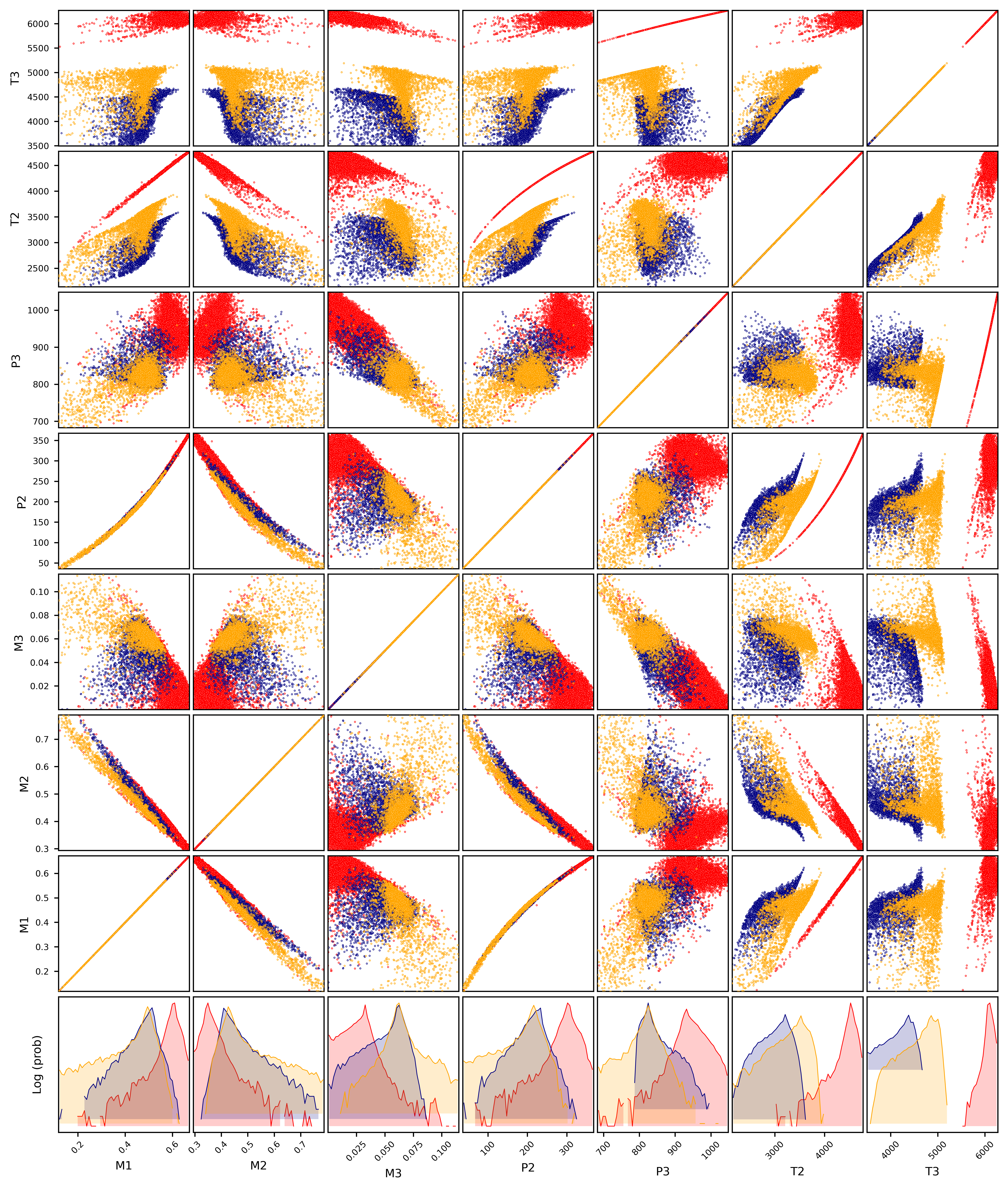}
\caption{Like in Fig.~\ref{fig:corr}, we show a posterior MCMC distribution but here we have reduced the number of variables and compare three distributions for Neptune: Red color represents models with the hotter temperature profile proposed by \citet{Redmer2011}. Models of type B with a lower temperature profile that assume a convective barrier are shown in blue. The orange results represent our C type models with intermediate temperatures that assume convection occurs across the superionic-liquid boundary.\label{fig:corr2}}
\end{figure}

To characterize the plausible range of model parameters, we
constructed ensembles of interior models with our MCMC methods
\citep{QMC_source_code,QMC_2024_source_code}. In Figs.~\ref{fig:Jn},
\ref{fig:corr}, and \ref{fig:corr2}, we show the resulting posterior
distributions for Uranus and Neptune models that we constructed with
the CMS method. The single flyby of the Voyager 2 spacecraft did not
constrain the gravity field of Neptune as well as that of
Uranus. The resulting distributions of the gravity harmonics $J_n$ in
Fig.~\ref{fig:Jn} are thus much wider for Neptune. These coefficients
describe a planet's response to rotation. For nonrotating planets in
hydrostatic equilibrium, all $J_{n>0}$ would be zero. In the
histograms, the $J_{2n}$ distributions of both planets overlap but in
the $J_{2n}-J_{2m}$ correlation plots, the Neptune models tend to show
larger $J_{2m}$ value, which is not too surprising because as Neptune
rotates a bit faster than Uranus. The signs of the $J_{2n}$ alternate
with increasing $n$, which is simply a consequence of how the Legendre
polynomials are defined. If one studies just the magnitudes of the
$J_{2n}$, one finds that they are all positively correlated with each
other in Figs.~\ref{fig:Jn}. So if according to one model in the MCMC
ensemble, the mass distribution inside a planet responds a bit more
strongly to the centrifugal forces, the magnitude of all $J_{2n}$
increases slightly, which explains their pairwise positive
correlation.

For all of our models, we assume Uranus and Neptune cool convectively
and their interior temperature profiles can be represented by
different isentropes. One still has to make additional assumptions for
the first-order transition from liquid to superionic water that is
associated with a discontinuous change in density and entropy as we
discussed in the previous section. If one assumes that this transition
introduces a barrier to convection then one would treat it like the
boundary between the hydrogen-helium and the ocean layers where one
assumes common values for pressure and temperature at the interface
but then choose a different value of the entropy in both layers. The
results are continuous pressure-temperature profiles as we illustrate
in Fig.~\ref{fig:models}. By assuming a continuous
pressure-temperature profile, \citet{Redmer2011} and
\citet{Nettelmann2013} have implicitly assumed that the transition to
superionic water introduces a convective barrier. It is therefore
useful discuss with such type of models. We refer to them as type
``B'' models.

In Fig.~\ref{fig:corr}, we compare the fractional layer masses and
pressure and temperatures at layer boundaries for different ensembles
of B-type models illustrated in Fig.~\ref{fig:models}. The blue and
red points respectively show models for Neptune that we derived with
our interior temperature profiles and those proposed by
\citet{Redmer2011} (R11). We compare with these profiles because they
are still referenced by experimentalists
\citep{prakapenka2021structure}. All results were obtained with the
CMS method under consistent assumptions. The black and orange dots
display results for Uranus with our and with R11's temperature
profile. Most interesting are panels in Fig.~\ref{fig:corr} where all
four distributions differ. An example is the correlation between the
temperatures $T_1$ and $T_2$, temperatures at the top and bottom of
the ocean layer. The four distributions vary because the planet
observations and the assumed isentropes differ. For our B-type models,
we predict the average values of $T_1$ and $T_2$ to be 1520 and 3000~K
for Neptune and 1890 and 3370~K for Uranus. If we construct models
under the same assumptions but adopt R11's temperature profiles, we obtain
1560 and 4520~K for Neptune's $T_1$ and $T_2$ and 1910 and 4510~K for
Uranus. The histograms at the bottom of Fig.~\ref{fig:corr} confirm
that our $T_2$ values for both planets are much lower than the
corresponding R11 values while $T_1$ are fairly similar because they
are not affected by our revision of the $T$-$P$ profile in the ocean
layer. The $T_1$ values for Uranus are higher simply because we
predict this planet to have a thick H-He layer that encompasses 4\% of
the planet's mass while we predict only 1.5\% for Neptune.

For $T_3$, the temperature between the C-N-H layer and the core, we
predict an average value of 4360 and 4120~K of Uranus and
Neptune. With R11's T-P profiles, we obtain much higher values of 6040
and 6100~K (also see Fig.~\ref{fig:H2O}). In comparison, the value of
the corresponding pressure, $P_3$ does not change much. We predict 620
and 820~GPa for Uranus and Neptune when we adopt our P-T profile and
630 and 940~GPa for the R11 profile. Similarly our predictions for
Neptune's $P_2$ (pressure at boundary of ocean and C-N-H layer) is
lower (average of 200 GPa) than values obtained with R11's temperature
profile (300 GPa). In general, Neptune's values are larger because it
is a bigger planet with a bigger core. For Neptune, we predict core mass fraction
of about 6.4\% (1.1 Earth masses) and a central pressure of 1300~GPa
while we predict a much smaller core or even no core for Uranus. The
average core mass fraction in our ensembles was only 0.17\% and the
central pressure was 650~GPa.

\begin{figure}[ht!]
\plotone{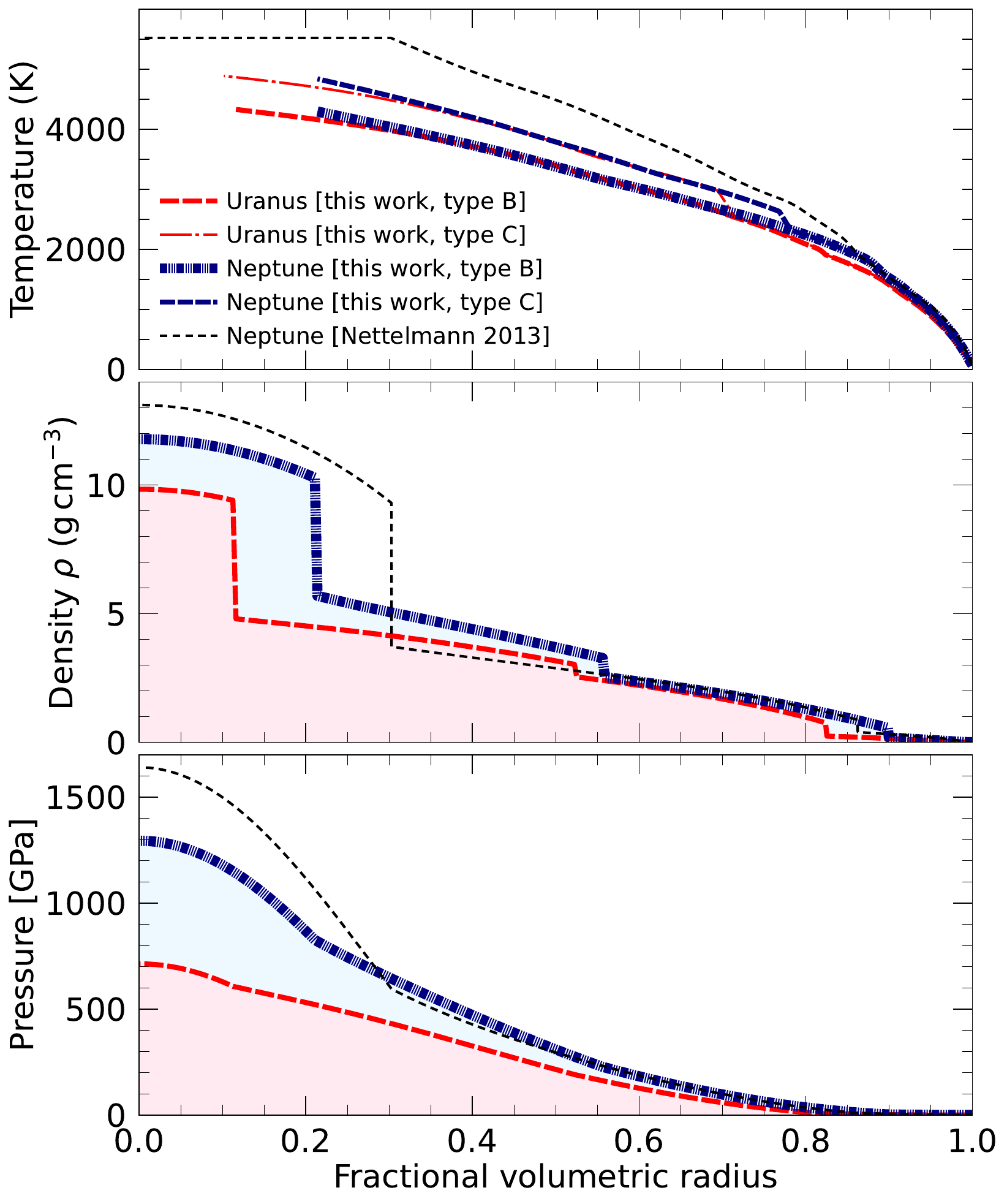}
\caption{Radial profiles of temperature, density, and pressure from our models in Fig.~\ref{fig:models}. For comparison, we show a model by \citet{Nettelmann2013} that proposed a much hotter interior and a larger core for Neptune.\label{fig:function_vs_R}}
\end{figure}

The $M_1$-$M_2$ panel of Fig.~\ref{fig:corr} shows that both mass
fraction are strongly anticorrelated because their densities are not
very different (see Fig.~\ref{fig:function_vs_R}) and one can thus
move their boundary up or down without drastically changing the
gravity field. For our P-T profile we predict the ocean and C-N-H
layers to encompass 57 and 39\% of Uranus and each 46\% of Neptune.

A number of planels in Fig.~\ref{fig:corr} show very strong
correlations. For example $T_1$ and $P_1$ correlated because they are
both functions of the mass of the H-He layer ($M_0$) and the outer
adiabats are anchored at their respective 1~bar temperatures of 72 or
76~K. Similarly one finds that $T_2$ and $P_2$ are positively
correlated with each other and with the mass of the ocean layer,
$M_1$. Not surprisingly, one finds the central pressure of both planets
to correlate with their core masses. One may also increase the core
mass by lowering the core transition pressure, $P_3$, which explain
why both quantities are anticorrelated.

Convection in the presence of phase transitions has been discussed in
context of geophysics because different mantle minerals undergo exothermic and
endothermic phase transitions in the Earth's mantle. With increasing
pressure or depth, the liquid-to-superionic transition of water is
similar to the exothermic transition of olivine to spinel that was
analyzed by \citet{schubert1975role}. When a cold slab descends,
olivine, the lower-pressure phase, transforms into spinel, a
higher-pressure polymorph. As its phase changes, it releases latent heat and its density
decreases. \citet{schubert1975role} discussed two competing
effects. First they argued that the release of latent heat would
increase the local temperature, which would lower the material's
density and introduce a buoyancy force that would counteract the
convecting forces. Second they pointed out that a descending slab must
have a temperature that is slightly lower than that of its surrounding,
which means it would encounter the olivine-to-spinel transition at
slightly lower pressures because the phase transition has a positive Clapeyron
slope. Therefore it would transition to the higher-density spinel
phase at a slightly shallower depth. There its density would be higher than
that of its surrounding, which would promote the convection across the
phase boundary.

\citet{christensen1995effects} analyzed the magnitude of the two
effects and concluded that for Rayleigh convection in Earth's mantle,
the second effect dominates over the first and that exothermic phase
transitions enhance convection in general. If one adopts this picture
for the convection of H$_2$O in the Uranus and Neptune then one is
forced to represent the liquid and superionic portions of their
mantles by two curves that have a same entropy. This would lead to an
offset in pressure-temperature space at the melting line of superionic
H$_2$O, which we illustrate with our models of type ``C'' in
Figs.~\ref{fig:H2O} and \ref{fig:models}. In the gap between liquid and superionic branches,
the isentrope would follow the melting line. In the planet, one would
then assume there exists a {\em transition zone} or a mixed region where
the fraction of superionic H$_2$O increases gradually with depth. If a
cold downwelling arrived at this transition zone, it would change to
superionic H$_2$O at slightly lower pressures, therefore become a bit
denser than its surrounding, and so continue to descend. The release
of latent heat would increase the temperature while the entropy
remains constant because heat diffusion into the environment remains
to be too slow. By similar arguments, an upwelling of hotter material
would also pass through the transition zone.

In Fig.~\ref{fig:corr2} we compare ensembles of models for Neptune
that have constructed by assuming a single entropy value but two
separate isentrope branches for the liquid and superionic phases. As
expected one observes an increase in the temperatures at the
water-to-CNH layer, $T_2$, of about 300~K (3400 instead of 3100~K) and
at the CNH-core boundary, $T_3$ of about 500~K (4750 instead of
4250~K). Both results are much colder than values of $T_2$=4500 and
$T_3$=6100~K that one would predict if one adopted the temperature
profile that was proposed by \citet{Redmer2011}. With this profile one
predicts smaller cores ($M_3$), less massive CNH layers ($M_2$) but
thicker water layers ($M_1$).

In Fig.~\ref{fig:corr2} also shows that for B and C type models, the
most likely values layer masses $M_1$, $M_2$, and $M_3$ and boundary
pressures $P_2$ and $P_3$ are very similar but the distribution of C
type models includes more massive cores that are accommodated by
smaller $P_2$ and $P_3$ values. At the present time, we have no
information to determine whether B and C type models more likely
represent the interiors of Uranus and Neptune, so we recommend
treating them as equally probable. 

In Fig.~\ref{fig:function_vs_R}, we show the radial profiles of
density, temperature and pressure for the presentative models in
Fig.~\ref{fig:models}. Because Neptune has a larger core than Uranus,
the pressure and density in its center are higher. As expected, the
temperature of our C-type models with a bit higher than that of our
B-type models but their density and pressure profiles are rather
similar. The temperature in a Neptune model by \citet{Nettelmann2013}
are significantly higher than ours, which explains why they predicted
a larger core for Neptune and the density in their water layer is
comparable to that in our water-hydrogen layer.

\section{Conclusions} \label{sec:conclusions} 

We performed {\it ab initio} computer simulations of liquid and
superionic H$_2$O for conditions of high pressure and temperature in
the interiors of Uranus and Neptune. We derived the free energies by
employing thermodynamic integration methods, which gave us a direct
access to the {\it ab initio} entropies. This enabled us to compute
isentropes and to construct models for the interiors of Uranus and
Neptune. We put together two types of models assuming that convection
may or may not reach across the transition between liquid and
superionic water.

We propose a revision for the interior temperature profiles of Uranus
and Neptune, as for both types of models we predict their deep
interiors to be much colder ($\sim$4300 or $\sim$4800) than earlier models by \citet{Redmer2011}
and \citet{Nettelmann2013} ($\sim$6000~K) who did not have access to entropies that were
computed with {\it ab initio} simulations. Models by
\citet{hubbard1980structure}, \citet{Chau2011}, and by
\citet{neuenschwander2024uranus} had predicted even higher
temperatures for the interiors of Uranus and Neptune, so consequently
our temperatures deviate from them even more.

Our predictions for lower temperatures in Uranus and Neptune have a
number of likely consequences. They make it much more likely for
diamonds to form \citep{ross1981ice} or for the phase separation of
planetary ices \citep{militzer2024phase} to have occurred in these
planets. Future compression experiments such as these performed by
\citet{Chau2011}, \citet{kraus2017formation}, and
\citet{frost2024diamond} may be designed to target a lower temperature
range. This means there is a smaller temperature differential
between the atmosphere and deep interiors, which implies that there is
less energy available to drive thermal convection and to generate
magnetic fields today. It also means that Uranus and Neptune have
radiated more thermal energy into space since their formation, which
evolution models will need to take into account when they target
today's interior state that is now substantially colder than was
previously predicted.

Based on results from our {\it ab initio} simulations, we constructed
static models for the interior structure of Uranus and Neptune with
the Concentric MacLaurin Spheroid method. Neptune is that predicted to
have a rocky core that comprised approximately one Earth mass while
Uranus is predicted to have a smaller rocky core or no core at all.

\begin{acknowledgments}
  Kyla de Villa provided comments on this manuscript. This work was
  supported by the Department of Energy-National Nuclear Security
  Administration (DE-NA0004147) via the Center for Matter at Extreme
  Conditions and by the Juno mission of the National Aeronautics and
  Space Administration.
\end{acknowledgments}

% \bibliography{UN}{}
% \bibliographystyle{aasjournal}

\end{document}